%% file: pure_tetrad_arxiv_v4.tex
\documentclass[
10pt, 
a4paper, 
oneside, 
twocolumn,
headinclude,footinclude, 
BCOR5mm, 
]{scrartcl}

\input{structure.tex} 

\usepackage[hmarginratio=1:1,top=25mm,left=17mm,columnsep=20pt]{geometry}
\usepackage{relsize} 
\usepackage{bm}

\newcommand{\ERWBB}{{ERWBB}}

\newcommand{\sA}{\mathsmaller A}
\newcommand{\sB}{\mathsmaller B}
\newcommand{\sC}{\mathsmaller C}
\newcommand{\sD}{\mathsmaller D}
\newcommand{\sE}{\mathsmaller E}

\newcommand{\pd}[1]{\partial_{#1}}

\newcommand{\mg}[1]{\kappa_{#1}}			
\newcommand{\MG}[1]{\kappa^{#1}}			

\newcommand{\pdd}[1]{{\bm{\partial}_{#1}}}
\newcommand{\dx}[1]{{\bm{\mathrm{d}x}^{#1}}}

\newcommand{\tetrsymbol}{h}
\newcommand{\itetrsymbol}{\eta}
\newcommand{\itetr}[2]{\itetrsymbol^{#1}_{\phantom{#1}#2}}
\newcommand{\tetr}[2]{\tetrsymbol^{#1}_{\phantom{#1}#2}}
\newcommand{\stress}[2]{s_{\ #1}^{#2}}

\newcommand{\detTetr}{\tetrsymbol}

\newcommand{\cobas}[1]{\bm{\tetrsymbol}^{#1}}
\newcommand{\bas}[1]{\bm{\itetrsymbol}_{#1}}

\newcommand{\D}[1]{\partial_{#1}} 
\newcommand{\DW}[1]{\mathcal{D}_{#1}} 
\newcommand{\Tors}[2]{T^{#1}_{\phantom{#1}#2}}

\newcommand{\ET}[2]{E^{#1}_{\phantom{#1}#2}}	
\newcommand{\ETmix}[2]{E^{#1}_{#2}}	
\newcommand{\Dm}[2]{D_{\phantom{#2}#1}^{#2}}	
\newcommand{\aD}[2]{\mathcal{D}_{\phantom{#2}#1}^{#2}}	
\newcommand{\Dfin}[2]{\mathtt{D}_{\phantom{#2}#1}^{#2}}	
\newcommand{\Hfin}[2]{\mathtt{H}_{#2#1}}	
\newcommand{\Efin}[2]{\mathtt{E}^{#1}_{\phantom{#1}#2}}	
\newcommand{\Ufin}{\mathtt{U}}
\newcommand{\Kbuch}[2]{\mathtt{K}_{{#1}{#2}}}	
\newcommand{\Nbuch}[2]{\mathtt{N}_{#1}^{\ \,#2}}	
\newcommand{\Nbuchdown}[2]{\mathtt{N}_{#1#2}}	
\newcommand{\Kbuchmix}[2]{\mathtt{K}^{#1}_{\ #2}}	
\newcommand{\Nbuchmix}[2]{\mathtt{N}^{#1}_{\ #2}}	
\newcommand{\BT}[2]{B^{#1#2}}	
\newcommand{\BTmix}[2]{B^{#1}_{#2}}	
\newcommand{\Bmmix}[2]{B^{#1}_{#2}}	
\newcommand{\Bm}[2]{B^{#1#2}}	
\newcommand{\aB}[2]{\mathcal{B}^{#1#2}}	
\newcommand{\Bfin}[2]{\mathtt{B}^{#1#2}}	
\newcommand{\Bfinmix}[2]{\mathtt{B}^{#1}_{\phantom{#1}#2}}	
\newcommand{\Bfinmixx}[2]{\mathtt{B}_{#1}^{\phantom{#1}#2}}	

\newcommand{\w}[2]{W^{#1}_{\phantom{#1}#2}}

\newcommand{\We}{Weitzenb\"ock}
\newcommand{\Lag}{\Lambda}	
\newcommand{\Laghodge}{L}
\newcommand{\Lagtors}{\mathfrak{L}}
\newcommand{\LagBE}{\mathcal{L}}
\newcommand{\Um}{U}
\newcommand{\aU}{\mathcal{U}}

\newcommand{\EMmat}[2]{\sigma^{#1}_{\ \,#2}}

\newcommand{\LCsymb}{\bm{\in}}    
\newcommand{\LCtens}{\varepsilon} 
\newcommand{\rhs}[1]{f_{#1}}
\newcommand{\mat}[1]{\prescript{\text{(m)}}{}{\hspace{-0.1cm}#1}}
\newcommand{\gra}[1]{\prescript{\text{(g)}}{}{\hspace{-0.1cm}#1}}

\newcommand{\tegr}{TEGR}
\newcommand{\HDT}[1]{\accentset{\star}{T}^{#1}}

\newcommand{\KD}[2]{\delta^{#1}_{\ #2}}
\newcommand{\NC}[2]{J^{#2}_{\phantom{#2}#1}}

\newcommand{\indalg}[1]{\hat{\mathsmaller{#1}}}

\newcommand{\TorsConj}[2]{\mathbb{T}_{#1}^{\phantom{#1}#2}}
\newcommand{\HTConj}[1]{\accentset{\star}{\mathbb{T}}_{#1}}
\newcommand{\Dbb}[2]{\mathbb{D}_{#1}^{\phantom{#1}#2}}
\newcommand{\Hbb}[2]{\mathbb{H}_{#1#2}}
\newcommand{\lapse}{\alpha}
\newcommand{\shift}[1]{\beta^{#1}}
\newcommand{\Tscal}{\mathcal{T}}		

\newcommand{\Hscal}{\mathcal{H}}		

\title{\large\normalfont\spacedallcaps{
		First-order 
		hyperbolic 
		formulation of the \\
		teleparallel gravity theory}} 

\author{\normalsize\textsc{Ilya Peshkov}$^{*,1}$,\ 
	\normalsize\textsc{Héctor Olivares}$^{2,3}$
	\& 
	\normalsize\textsc{Evgeniy Romenski}$^{4}$
} 

\date{\small\today} 

\begin{document}
	
	
	\renewcommand{\sectionmark}[1]{\markright{\spacedlowsmallcaps{#1}}} 
	\lehead{\mbox{\llap{\small\thepage\kern1em\color{halfgray} 
				\vline}\color{halfgray}\hspace{0.5em}\rightmark\hfil}} 
	
	\pagestyle{scrheadings} 
	
	
	\maketitle 
	
	\setcounter{tocdepth}{2} 
	
	
	
	
	
	\section*{Abstract} 
	\noindent
	\textit{ This paper presents a derivation of a first-order reduction and 3+1
	decomposition of the teleparallel equivalent of general relativity (TEGR) in
	the pure-tetrad formulation (no spin connection). Our analysis demonstrates
	that in vacuum spacetimes, our 3+1 TEGR equations has the principal part of
	the differential operator equivalent to the one of tetrad reformulation of
	general relativity by Estabrook, Robinson, Wahlquist, and Buchman and
	Bardeen, and therefore the presented 3+1 decomposition of TEGR also admits a
	symmetric hyperbolic formulation, a desirable property for ensuring
	well-posedness of the initial value problem. Furthermore, the structure of
	the 3+1 equations possess a lot of similarities with the equations of
	relativistic electrodynamics and the recently proposed dGREM
	tetrad-reformulation of general relativity. }
	\footnotetext{* Corresponding author, \textit{ilya.peshkov@unitn.it}}
	\footnotetext{\textsuperscript{1} \textit{Department of Civil, Environmental and Mechanical 
			Engineering, University of Trento, Via
			Mesiano 77, 38123, Trento, 
			Italy}}
	\footnotetext{\textsuperscript{2} \textit{
			Department of Astrophysics/IMAPP, Radboud University Nijmegen, P.O. Box 9010, NL-6500 
			GL Nijmegen, 
			Netherlands}}
	\footnotetext{\textsuperscript{3} \textit{
			Center for Research and Development in Mathematics and Applications (CIDMA), 
			Department of Mathematics, University of Aveiro, 3810-193 Aveiro, Portugal}}
	\footnotetext{\textsuperscript{4} \textit{Sobolev Institute of Mathematics, Novosibirsk, 
	Russia}}
	\renewcommand{\thefootnote}{\arabic{footnote}}
	
	
	\setlength\parindent{10pt} 
	\setlength{\parskip}{5pt} 
	
	
	\section{Introduction}
	
	The class of teleparallel gravity theories is one of the alternative
	reformulations of Einstein's general relativity (GR)
	\cite{Hehl1976,AldrovandiPereiraBook,Cai2016}. While for GR gravitational
	interaction is a manifestation of the curvature of a torsion-free spacetime,
	for the teleparallel framework it is realized as a curvature-free linear
	connection with non-zero torsion (or/and non-zero nonmetricity which is not
	considered in this paper, e.g. \cite{Adak2006}). Although different variables can be taken as the
	main dynamical fields in GR (tetrad fields, soldering forms, etc.), the most
	extended choice is the metric tensor accompanied by the Levi-Civita
	connection. In contrast, for teleparallel theory the metric is trivial and
	the main dynamical field is usually the space-time tetrad (or frame) field.
	
	Despite the teleparallel geometries are considered as an alternative
	framework\footnote{Yet, it is guaranteed that simplest realizations of
	teleparallel geometries, such as TEGR for example, produce all the classical
	results of general relativity \cite{AldrovandiPereiraBook,Bahamonde2021a}.}
	to Einstein's gravity with several promising features missing in GR, e.g.
	see the discussion in \cite[Sec.18]{AldrovandiPereiraBook} and
	\cite{Cai2016}, in this paper, we are interested in the teleparallel gravity
	only from a pure computational viewpoint and our goal is to use its
	mathematical structure to develop an efficient computational framework for
	numerical relativity. Thus, the main goal of this paper is to derive a
	3+1-split of the so-called \emph{teleparallel equivalent of general
	relativity} (TEGR) \cite{AldrovandiPereiraBook,Krssak2019} which is known to
	pass all standard tests of GR. Up to now, not many attempts have been done
	to obtain a $ 3+1 $ formulation of the TEGR equations, e.g.
	\cite{Maluf2001a,Capozziello2021,Pati2022}. A Hamiltonian formulation of
	TEGR was used in \cite{Maluf2001a,Pati2022} to obtain evolution equations
	for the tetrads and conjugate momenta. In \cite{Capozziello2021}, the
	spatial tetrad and their first order Lie derivative along the normal vector
	to the foliations were chosen as the state variables. Somewhat similar to
	the Hamiltonian formalism in systematic use of conjugate variables and
	Legendre transforms, here we explore another line of deriving a 3+1
	formulation of TEGR that is aligned with the relativistic
	electrodynamics. In particular, the key difference from the Hamiltonian
	formulations \cite{Maluf2001a,Pati2022} is in the promotion of the torsion
	to the dynamical variable with its own evolution equation. As the result,
	the obtained 3+1 equations are completely different from the mentioned
	papers and have a rather elegant structure similar to Maxwell's equations.

	A second objective of this paper is to develop a 3+1 formulation entirely
	independent of the Einstein's theory, i.e. our development avoids
	foundational GR tools like the Levi-Civita connection and the
	Hilbert-Einstein action. Instead, we start from an arbitrary Lagrangian
	density which is a function of the torsion scalar and obtain corresponding
	Euler-Lagrange equations coupled with some constraints (differential
	identities). The obtained equations are then reduced to the first-order form
	and a 3+1 decomposition is performed. Although a completely different route
	has been taken to obtain our 3+1 TEGR equations, we demonstrate that the
	resulting equations are equivalent to the recently proposed tetrad
	reformulation of GR, called dGREM, \cite{Olivares2022}.

	From the numerical view point, any 3+1 formulation has to have the
	well-posed initial value problem (i.e. a solution exists, the solution is
	unique and changes continuously with changes in the initial data) in order
	to compute stable evolution of the numerical solution. In other words, the
	system of governing equations has to be hyperbolic\footnote{Note that in the
	numerical relativity the term ``strong hyperbolicity'' is used emphasizing
	that not only eigenvalues must be real but that the full basis of
	eigenvectors must exist.}. Therefore, the third objective of the paper is to
	test if the proposed 3+1 formulation of TEGR is hyperbolic. As is the case
	with other first-order reductions of the Einstein equations
	\cite{Baumgarte2003a}, the question of hyperbolicity of 3+1 TEGR equations
	considered here is not trivial and depends on the delicate use of multiple
	involution constraints (stationary identities), e.g. see \cite{FO-CCZ4}. In
	particular, for vacuum spacetimes and a certain choice of gauge conditions,
	we have found that the proposed 3+1 TEGR equations can be transformed into a
	symmetric hyperbolic system similar to the tetrad reformulation of the GR
	proposed by Estabrook, Robinson, and Wahlquist \cite{Estabrook1997} and
	further developed by Buchman and Bardeen
	\cite{Buchman2003,Buchman2005,Bardeen2011} (the \ERWBB\ formulation for
	brevity). 
	
	We note that we do not consider the most general TEGR formulation as for
	example presented in \cite{AldrovandiPereiraBook}. The linear connection of
	TEGR is the sum of the two parts: the \We\ connection (which is the
	historical connection of TEGR) and the spin connection (parametrized by
	Lorentz matrices) representing the inertial content of the tetrad. The spin
	connection is necessary to separate the inertial effect of a chosen frame
	from its gravity content as well as for establishing the full covariance of
	the theory \cite{AldrovandiPereiraBook,Golovnev2017a,Krssak2019}, i.e. with
	respect to both the diffeomorphisms of the spacetime and the Lorentz
	transformations of tangent spaces. However, being important from the
	theoretical viewpoint and for extensions of teleparallel gravity
	\cite{Golovnev2017a}, the spin connection introduces extra degrees of
	freedom which do not have evolution equations in TEGR and therefore can be
	treated as non-dynamical parameters of the theory. Since we are
	interested in developing a computational framework for GR, the consideration
	of this paper is, therefore, restricted to the frames for which the spin
	connection is set to zero globally (\We\ gauge). We thus consider TEGR in
	its historical, or \emph{pure tetrad}, formulation \cite{Golovnev2017a}.

	Another motivation to study the mathematical structure of
	the teleparallel gravity is coming from the continuum fluid and solid
	mechanics. In particular, the role of the torsion to describe defects in
	solids has been known for decades now, e.g. see
	\cite{VolovichKatanaev1992,Hehl2007,Yavari2012,NguyenLeMarrec2022,Bohmer2020,Lychev2022}.
	Moreover, the material tetrad field (called also the distortion field in our
	papers) is the key field for the unified hyperbolic model of fluid and solid
	mechanics \cite{HPR2016,DPRZ2016}. In such a theory, the concept of torsion
	can be connected with the inertial effect of small-scale eddies in turbulent
	flows and with dispersion effects in heterogeneous solids (e.g. acoustic
	metamaterials) as discussed in \cite{Torsion2019}. Interestingly, that the
	3+1 equations we obtain in this paper resemble very closely the structure
	of the equations for continuum fluid and solid mechanics with torsion
	\cite{Torsion2019}. Furthermore, the unified theory of fluids and solids has
	been also extended in the general relativistic settings \cite{PTRSA2020} and
	therefore, as being a tetrad theory by its nature, it can be
	straightforwardly coupled with the 3+1 TEGR equations discussed in this
	paper.

	This work is organized as follows. We start with recalling important
	definitions in Section\,\ref{sec.def}. In Section\,\ref{sec.SAP}, we recall
	the Euler-Lagrange equations for a general Lagrangian defined as an
	arbitrary function of the tetrad field and its first gradients. Then, in
	Section\,\ref{sec.PDEs}, we replace the second-order Euler-Lagrange
	equations by an extended system of first-order partial differential
	equations. In Section\,\ref{sec.31.prep}, we rewrite the obtained system in
	a form similar to relativistic electrodynamics. Section \ref{sec.31}
	discusses the details of the 3+1 split of the TEGR equations, which is the
	main result of this paper. The summary of the 3+1 equations is given in
	Section\,\ref{sec.summary}. In Section\,\ref{sec.hyperbolicity}, we discuss
	the question of hyperbolicity of the vacuum 3+1 TEGR equations via
	transforming these equations into an equivalent symmetric hyperbolic system similar to that of \ERWBB\ \cite{Estabrook1997,Buchman2003}. Finally, we
	outline possible directions for future research in
	Section\,\ref{sec.conclusion}.

	\section{Definitions}\label{sec.def}
	
	\subsection{Non-holonomic frame field}
	
	Throughout this paper, we use the following index convention. Greek letters $ \alpha, 
	\beta\,\gamma$, 
	$\ldots, \lambda,\mu,\nu,... 
	=0,1,2,3
	$ are used to index quantities related to the spacetime manifold, the Latin letters from the 
	beginning of the alphabet $ a,b,c,... 
	=\hat{0},\hat{1},\hat{2},\hat{3}$ are used to index quantities related to the tangent Minkowski 
	space.
	In the 3+1 split, the letters $ i,j,k,\ldots =1,2,3$ from the middle of the Latin alphabet 
	are 
	used 
	to denote spatial components of the spacetime tensors, and the upper case Latin letters $ 
	\sA,\sB,\sC,\ldots=\indalg{1},\indalg{2},\indalg{3} $ index spatial components of the tensors 
	written in a chosen frame of the 
	tangent space.

	We shall consider a spacetime manifold $ M $ equipped with a general Riemannian metric $ 
	g_{\mu\nu} 
	$ and a coordinate system $ x^\mu $. At each 
	point of 
	the spacetime, there is the tangent space $ T_{x}M $ spanned by the frame (or tetrad) $ 
	\pdd{\mu} $ which is the standard coordinate basis. 
	There is also the cotangent space $ T_x^*M $ spanned by the coframe field $ \dx{\mu} $. 
	Recall that the frames $ \pdd{\mu} $ and $ \dx{\mu} $ are \emph{holonomic} 
	\cite{AldrovandiPereiraBook}.
	
	
	In addition to $ T_{x}M $, we assume that at each point of $ M $, there is a soldered tangent 
	space 
	which is a Minkowski space spanned by an orthonormal frame (tetrad) $ \bas{a} $ and equipped 
	with 
	the 
	Minkowski metric 
	\begin{equation}\label{eqn.mg}
		\mg{ab} = \text{diag}(-1,1,1,1).
	\end{equation}
	Similarly, there is the 
	corresponding cotangent space spanned by the co-frames $ \cobas{a} $. It is assumed 
	that the frames $ \cobas{a} $ and $ \bas{a} $ are independent of the coordinates $ x^\mu $ and, 
	therefore, are non-holonomic in general.
	
	The components of the non-holonomic frames $ \cobas{a} $ and $ \bas{a} $ in the holonomic 
	frames $ \dx{\mu} $ and $ \pdd{\mu} $ are denoted by $ \tetr{a}{\mu} $ and $ \itetr{\mu}{a} $, 
	i.e. 
	\begin{equation}
		\cobas{a} = \tetr{a}{\mu}\dx{\mu}, \qquad \text{or} \qquad \bas{a} = \itetr{\mu}{a}\pdd{\mu}
	\end{equation}
	with $ \tetr{a}{\mu} $ being the inverse of $ \itetr{\mu}{a} $, i.e.
	\begin{equation}\label{eqn.inv.tetr}
		\tetr{a}{\mu} \itetr{\mu}{b} = \KD{a}{b},
		\qquad
		\itetr{\nu}{a}\tetr{a}{\mu}  = \KD{\nu}{\mu},
	\end{equation}
	where $ \KD{a}{b} $ and $ \KD{\nu}{\mu} $ are the Kronecker deltas.
	
	The soldering of the spacetime and the tangent Minkowski space means that the metrics $ 
	g_{\mu\nu} 
	$ and $ \mg{ab} $ are related by  
	\begin{equation}
		g_{\mu\nu} = \mg{ab} \tetr{a}{\mu}\tetr{b}{\nu}.
	\end{equation}
	
	
	Note that 
	\begin{equation}\label{eqn.det}
		\detTetr := \det(\tetr{a}{\mu}) = \sqrt{-g},
	\end{equation}
	if $ g = \det(g_{\mu\nu}) $.
	
	\subsection{Observer's 4-velocity}
	
	We choose to associate observer's 4-velocity with the $ 0 $-\textit{th} vector of the tetrad basis 
	\begin{equation}\label{eqn.4v}
		u^\mu := \itetr{\mu}{\indalg{0}}, \qquad u_\mu = g_{\mu\nu}u^\nu.
	\end{equation}
	Also, due to
	\begin{equation}
		u_\mu = g_{\mu\nu} u^\nu = \mg{ab}\tetr{a}{\mu}\tetr{b}{\nu}\itetr{\nu}{\indalg{0}} = 
		\mg{a\indalg{0}}\tetr{a}{\mu} = -\tetr{\indalg{0}}{\mu},
	\end{equation}\label{eqn.4v.cov}
	the covariant components of the 4-velocity equal to entries of the $ 0 $-\textit{th} vector of 
	the 
	co-basis with the opposite sign
	\begin{equation}
		u_\mu = -\tetr{\indalg{0}}{\mu}.
	\end{equation}
	%

	\subsection{Connection and torsion}
	
	Because we work in the framework of the pure-tetrad formulation of TEGR
	(\We\ gauge), the linear connection is set to the pure \We\
	connection\footnote{Note that we use a different convention on the
	positioning of the lower indices of the \We\ connection $ \w{a}{\mu\nu} =
	\pd{\mu}\tetr{a}{\nu} $ than in \cite{AldrovandiPereiraBook}. Precisely, the
	derivative index goes first.}
	\cite{AldrovandiPereiraBook,KleinertMultivalued}: 
	\begin{equation}\label{eqn.We}
		\w{a}{\mu\nu} := \pd{\mu}\tetr{a}{\nu}, 
		\qquad
		\text{or}
		\qquad
		\w{\lambda}{\mu\nu} := \itetr{\lambda}{a}\pd{\mu}\tetr{a}{\nu}.
	\end{equation}
	The torsion is then defined as
	\begin{equation}\label{eqn.def.tors}
		\Tors{a}{\mu\nu}:=\D{\mu}\tetr{a}{\nu} - \D{\nu}\tetr{a}{\mu} = 
		\w{a}{\mu\nu} - \w{a}{\nu\mu}.
	\end{equation}
	
	Note that while the spacetime derivatives commute
	\begin{align}\label{eqn.commut.D}
		\D{\mu}(\D{\nu} V^\lambda) - \D{\nu}(\D{\mu} V^\lambda) &= 0, 
		\\[2mm] 
		\D{\mu}(\D{\nu} V^a) - \D{\nu}(\D{\mu} V^a) &= 0,
	\end{align}
	their tangent space counterparts $\D{a} =  \itetr{\mu}{a}\D{\mu}$ do not (for non-vanishing 
	torsion)
	\begin{equation}
		\D{b}(\D{c} V^a) - \D{c}(\D{b} V^a) = 
		-\Tors{d}{b c}\D{d}V^a,
	\end{equation}
	where $  \Tors{d}{bc} = \Tors{d}{\mu\nu}\itetr{\mu}{b}\itetr{\nu}{c} $.
	
	%

	\subsection{Levi-Civita symbol (tensor density)}
	
	
	We shall also need the Levi-Civita symbol (tensor-density\footnote{We 
		use the 
		sign convention for the tensor density weight according to \cite{Ryder2009,Grinfeld2013}, 
		i.e.
		under a 
		general 
		coordinate change $ x^\mu \to x^{\mu'} $, the determinant $ \det(\tetr{a}{\mu}) = \detTetr $ 
		transforms as $ \detTetr' = \det \left(\frac{\partial x^\mu}{\partial x^{\mu'}} \right)^W 
		\cdot \detTetr $ with $ W=+1 $. Therefore, the tetrad's determinant $ \detTetr $ and the 
		square root of the metric determinant $ \detTetr = \sqrt{-g} $ have weights $ +1 $, as well 
		as 
		the Lagrangian density in the action integral.} of weight $ +1 $)
	\begin{equation}\label{eqn.LCsymbol.def}
		\LCsymb^{\lambda\mu\nu\rho} = 
		\left\{ 
		\begin{array}{ll}
			+1,	& \text{if \ }\lambda\mu\nu\rho \text{ is an even permutation of } 0123,\\[2mm]
			-1,	& \text{if \ }\lambda\mu\nu\rho \text{ is an odd \ permutation of } 0123,\\[2mm]
			\phantom{-}0,	& \text{otehrwise}.
		\end{array}
		\right.
	\end{equation}
	Its covariant components $ \LCsymb_{\lambda\mu\nu\rho} $ define a tensor density of weight $ -1 
	$ 
	with the reference value $ \LCsymb_{0123} = -1 $. One could define an absolute 
	Levi-Civita 
	contravariant $ \LCtens^{\lambda\mu\nu\rho} = h^{-1} \LCsymb^{\lambda\mu\nu\rho} $ 
	and covariant $ \LCtens_{\lambda\mu\nu\rho} = h \LCsymb_{\lambda\mu\nu\rho} $ ordinary tensors  
	but for our further considerations (see Sec.\,\ref{sec.PDEs}), it is important that 
	the derivatives $ \D{\sigma}\LCsymb^{\lambda\mu\nu\rho} $ vanish:
	\begin{equation}\label{eqn.diff.LCsymb}
		\D{\sigma}\LCsymb^{\lambda\mu\nu\rho} = 0,
	\end{equation}
	whereas for $ \LCtens^{\lambda\mu\nu\rho} $ one has
	\begin{multline}\label{eqn.diff.LeviCivita}
		\D{\sigma}\LCtens^{\lambda\mu\nu\rho} = 
		\pd{\sigma}(\detTetr^{-1}\LCsymb^{\lambda\mu\nu\rho}) = 
		\LCsymb^{\lambda\mu\nu\rho}\pd{\sigma}\detTetr^{-1}   = \\[2mm] 
		-\LCsymb^{\lambda\mu\nu\rho}\detTetr^{-1}\itetr{\eta}{a}\pd{\sigma}\tetr{a}{\eta} = 
		-\LCtens^{\lambda\mu\nu\rho}\w{\eta}{\sigma\eta},
	\end{multline}
	which is not zero in general.

	%

	\section{Variational formulation}\label{sec.SAP}
	
	We consider a general Lagrangian (scalar-density) $ 
	\Lag(\tetr{a}{\mu},\pd{\lambda}\tetr{a}{\nu}) $ 
	of the teleparallel gravity which is a function of the frame field $ \tetr{a}{\mu} $ and its 
	first 
	gradients $ \w{a}{\lambda\nu} = \pd{\lambda}\tetr{a}{\nu} $. In what follows, we shall not 
	explicitly split $ \Lag $ 
	into the gravity (g) and matter (m) parts (unless it is explicitly mentioned otherwise), i.e. 
	\begin{equation}\label{eqn.Lagr.split}
		\Lag = \Lag^\text{(m)}(\tetr{a}{\mu},\pd{\lambda}\tetr{a}{\mu},\ldots) + 
		\Lag^\text{(g)}(\tetr{a}{\mu},\pd{\lambda}\tetr{a}{\nu}) 
	\end{equation}
	but the derivation will be performed for the total unspecified Lagrangian $
	\Lag $, so that in principle some elements of our derivation can be adopted
	for the extensions of the teleparallel gravity such as $ f(\Tscal)
	$-teleparallel gravity \cite{li2011d,li2011e,ferraro2018,blagojevic2020}, in
	which the Lagrangian $ \Lag = f(\Tscal) $ is defined as some function of the
	torsion scalar $ \Tscal $, see Section\,\ref{sec.closure}. We shall utilize
	the explicit form of the TEGR Lagrangian only in the last part of the paper,
	Section\,\ref{sec.31}. Yet, we emphasize that the $f(\Tscal)$-type
	teleparallel theories remain beyond the scope of this paper. 
	
	Varying the action of the teleparallel gravity $ \int 
	\Lag(\tetr{a}{\mu},\pd{\lambda}\tetr{a}{\mu}) 
	\bm{\dx{}}$ 
	with respect to 
	the tetrad, one obtains the Euler-Lagrange equations 
	\begin{equation}\label{eqn.EL}
		\frac{\delta \Lambda}{\delta\tetr{a}{\mu}} = \pd{\lambda}(\Lag_{\pd{\lambda}\tetr{a}{\mu}}) 
		- 
		\Lag_{\tetr{a}{\mu}} = 0,
	\end{equation}
	where $ \Lag_{\pd{\lambda}\tetr{a}{\mu}} = \frac{\partial 
		\Lag}{\partial(\pd{\lambda}\tetr{a}{\mu})} $ and $ 
	\Lag_{\tetr{a}{\mu}} = \frac{\partial \Lag}{\partial\tetr{a}{\mu}} $. Equations \eqref{eqn.EL} 
	form a system of 16 \emph{second-order} partial differential equations for 
	16 unknowns $ \tetr{a}{\mu} $. Our goal is to replace this second-order system by an equivalent 
	but 
	larger system of only first-order partial differential equations.
	
	\section{Equivalence to GR}
	
	Before writing system \eqref{eqn.EL} as a system of first-order equations let us first make a 
	comment on the equivalence of the GR and TEGR formulations. 
	
	The Lagrangian density of TEGR (and its extensions) is formed from the torsion scalar $ \Tscal 
	$, see \eqref{eqn.TEGR.Lagr}. As it is known, e.g. see \cite[Eq.(9.30)]{AldrovandiPereiraBook}, 
	the 
	torsion scalar can be written as
	\begin{equation}\label{eqn.TR}
		h \Tscal = -\sqrt{-g} R - \pd{\mu}(2 h \Tors{\nu}{\lambda\nu}g^{\lambda
			\mu}),
	\end{equation}
	where $ R $ is the Ricci scalar and $ \Tors{\nu}{\lambda\mu} = \itetr{\nu}{a} 
	\Tors{a}{\lambda\mu} 
	$. In other words, the Lagrangians of TEGR and GR differ by the four-divergence term (surface 
	term). The latter does not affect the Euler-Lagrange equations if there are no boundaries which 
	is 
	implied in this paper. 
	Therefore, Euler-Lagrange equations of TEGR \eqref{eqn.EL} in vacuum is nothing else but the 
	Euler-Lagrange 
	equations of GR written in terms of the tetrads and hence, their physical solutions must be 
	equivalent because the information about the physical interaction is contained in the 
	Euler-Lagrange equations of a theory. What is different in GR and TEGR is the 
	way one interprets the tetrads and their first derivatives, i.e. the way one defines the linear 
	connection of the spacetime from the gradients of tetrads, e.g. torsion-free Levi-Civita 
	connection 
	of GR and curvature-free \We\ connection of TEGR. These different geometrical 
	interpretations then define extra evolution equations (\emph{compatibility 
	constraints/identities}, 
	e.g. see \eqref{integr.HT}) 
	that are merely consequences of the geometrical definitions but do not define the physics of 
	the 
	gravitational interaction. 
	The critical point for the numerical relativity, though, is that these extra evolution 
	equations 
	must be solved simultaneously with the Euler-Lagrange equations and may affect the mathematical 
	regularity (well-posedness of the Cauchy problem) of the resulting system. 
	
	
	%

	\section{First-order extension}\label{sec.PDEs}
	
	Our first goal is to replace second-order system \eqref{eqn.EL} by a larger but first-order 
	system. 
	This is achieved in this section.
	
	From now on, we shall treat the frame field $ \tetr{a}{\mu} $ and its 
	gradients (the 
	\We\ 
	connection) $ 
	\pd{\lambda}\tetr{a}{\mu} $ formally as independent variables and in what 
	follows, we shall rewrite 
	system 
	of second-order PDEs \eqref{eqn.EL} as a larger system of first-order PDEs for the extended set 
	of  
	unknowns $ \{ \tetr{a}{\mu},\pd{\lambda}\tetr{a}{\mu} \} $, or actually, for their equivalents.
	
	In the setting of teleparallel gravity, $ \Lag $ is not a function of a general combination 
	of 
	the gradients $ \pd{\lambda}\tetr{a}{\nu} $ but of their special combination, that is torsion. 
	Yet, 
	we shall employ not the torsion directly but its Hodge dual, i.e. we assume that
	
	\begin{equation}\label{eqn.Lagrangians}
		\Lag(\tetr{a}{\mu},\pd{\lambda}\tetr{a}{\nu}) = 
		\Laghodge(\tetr{a}{\mu},\HDT{a\mu\nu}),
	\end{equation}
	where $ \HDT{a\mu\nu} $ is the Hodge dual to the 
	torsion, i.e.
	\begin{subequations}
		\begin{align}\label{eqn.Hodge.def}
			\HDT{a\mu\nu} &:= \frac{1}{2}\LCsymb^{\mu\nu\rho\sigma}\Tors{a}{\rho\sigma} = 
			\LCsymb^{\mu\nu\rho\sigma}\D{\rho}\tetr{a}{\sigma}, \\[2mm] 
			\Tors{a}{\mu\nu} &= 
			-\frac{1}{2}\LCsymb_{\mu\nu\rho\sigma}\HDT{a\rho\sigma}.
		\end{align}	
	\end{subequations}
	It is important to emphasize that we deliberately chose to define the Hodge dual using the 
	Levi-Civita symbol $ 
	\LCsymb^{\lambda\mu\nu\rho} $ and \emph{not} the Levi-Civita tensor $ 
	\LCtens^{\lambda\mu\nu\rho} = 
	\detTetr^{-1} 
	\LCsymb^{\lambda\mu\nu\rho} $ that will be important later for 
	the so-called integrability condition \eqref{integr.HT}.
	Remark that according to definition \eqref{eqn.Hodge.def}, $ \HDT{a\mu\nu} $ is a  
	\emph{tensor-density} of weight $ +1 $.

	In terms of the Lagrangian density $ \Laghodge(\tetr{a}{\mu},\HDT{a\mu\nu}) 
	$, using notations 
	\eqref{eqn.Lagrangians} and definitions 
	\eqref{eqn.Hodge.def}, we can instead
	rewrite Euler-Lagrange equations \eqref{eqn.EL} as
	\begin{equation}\label{eqn.EM.Hodge}
		\D{\nu}(\LCsymb^{\mu\nu\lambda\rho}\Laghodge_{\HDT{a\lambda\rho}}) 
		=-\Laghodge_{\tetr{a}{\mu}}.
	\end{equation}
	
	The latter has to be supplemented by the integrability 
	condition
	\begin{equation}\label{integr.HT}
		\D{\nu}\HDT{a\mu\nu} = 0,
	\end{equation}
	which is a trivial consequence of the definition of the  Hodge dual \eqref{eqn.Hodge.def}, i.e. 
	of 
	the 
	commutativity property of the standard spacetime derivative $ \D{\mu} $, and the 
	identity \eqref{eqn.diff.LCsymb}.
	We note that if the Hodge dual was defined using the Levi-Civita tensor $ 
	\LCtens^{\mu\nu\rho\sigma} $ instead of the Levi-Civita symbol, then one would 
	have that $ \D{\mu}\HDT{a\mu\nu} \neq 0 $.
	
	Another consequence of the commutative property of $ \pd{\mu} $ and the definition of the Hodge 
	dual (based on the Levi-Civita symbol) is that the Noether energy-momentum 
	current density
	\begin{equation}\label{eqn.Noether.current}
		\NC{a}{\mu} := \Laghodge_{\tetr{a}{\mu}}
	\end{equation}
	is conserved in the ordinary sense:
	\begin{equation}\label{eqn.Noether.cons}
		\D{\mu} \NC{a}{\mu} = 0.
	\end{equation}
	If  equations \eqref{eqn.EM.Hodge}, \eqref{integr.HT} and \eqref{eqn.Noether.cons} are 
	accompanied 
	with the 
	torsion definition
	\begin{subequations}
		\begin{align}\label{eqn.tetr}
			\D{\mu}\tetr{a}{\nu} - \D{\nu}\tetr{a}{\mu} &= \Tors{a}{\mu\nu},
			\qquad
		\end{align}	
	\end{subequations}
	they form the following system of \emph{first-order} partial differential
	equations (only first-order derivatives are involved)
	\begin{subequations}\label{eqn.1st.order.TEGR}
		\begin{align}	
			\D{\nu}(\LCsymb^{\mu\nu\lambda\rho}\Laghodge_{\HDT{a\lambda\rho}}) 
			&=-\Laghodge_{\tetr{a}{\mu}},\label{eqn.TEGR0.EL}\\[2mm]
			\D{\nu}\HDT{a\mu\nu} & = 0,\label{eqn.TEGR0.integr}\\[2mm]
			\D{\mu}\Laghodge_{\tetr{a}{\mu}} & = 0,\label{eqn.TEGR0.enermomen}\\[2mm]
			\D{\mu}\tetr{a}{\nu} - \D{\nu}\tetr{a}{\mu} &= \Tors{a}{\mu\nu},\label{eqn.TEGR0.tetrad}
		\end{align}
	\end{subequations}
	for the unknowns $ \{\tetr{a}{\mu},\HDT{a\mu\nu}\} $.
	
	This system forms a base on which we shall build our 3+1-split of TEGR in 
	Sections\,\ref{sec.31.prep} and \ref{sec.31}.

	
	\section{Energy-momentum balance laws}
	
	Any conservation law written as a  4-\emph{ordinary} divergence is a
	true conservation law, meaning that it yields a time-conserved 
	``charge'' \cite{AldrovandiPereiraBook}. Hence, the Noether current $ \NC{a}{\mu} 
	=\Laghodge_{\tetr{a}{\mu}}$ is a conserved charge in the ordinary sense, see 
	\eqref{eqn.Noether.cons}, \eqref{eqn.TEGR0.enermomen}. It expresses the conservation of the 
	total\footnote{The ``total'' here 
		means the gravitational + matter/electromagnetic energy-momentum current, i.e. $ 
		\NC{\mu}{a} = \Laghodge^\text{(g)}_{\tetr{a}{\mu}} + \Laghodge^\text{(m)}_{\tetr{a}{\mu}} 
		$.} 
	energy-momentum current density. 
	
	However, its spacetime counterpart 
	\begin{equation}\label{eqn.stress}
		\EMmat{\mu}{\nu}:=\tetr{a}{\nu} L_{\tetr{a}{\mu}},
	\end{equation}
	which can be called the total energy-momentum tensor density, is not
	conserved neither in the ordinary sense nor in the covariant one.
	
	Indeed, after contracting with $ \tetr{a}{\nu} $ and adding to it $ 0\equiv
	\Laghodge_{\tetr{a}{\mu}}\pd{\mu} \tetr{a}{\nu} -
	\Laghodge_{\tetr{a}{\mu}}\pd{\mu} \tetr{a}{\nu}  =
	\Laghodge_{\tetr{a}{\mu}}\pd{\mu} \tetr{a}{\nu} - \Laghodge_{\tetr{a}{\mu}}
	\tetr{a}{\lambda}\w{\lambda}{\mu\nu} $, Noether current conservation law
	\eqref{eqn.Noether.cons} can be rewritten in a pure spacetime form:	
	\begin{equation}\label{eqn.EM2}
		\pd{\mu}\EMmat{\mu}{\nu} = \EMmat{\mu}{\lambda} 
		\w{\lambda}{\mu\nu},
	\end{equation}
	which has a production term on the right hand-side and, therefore, $
	\EMmat{\mu}{\nu} $ is not a conserved quantity in the ordinary sense.
	Both forms \eqref{eqn.TEGR0.enermomen} and \eqref{eqn.EM2} will be put into a 3+1 form in Section \ref{sec.31}.

	On the other hand, if 
	\begin{subequations}
		\begin{align}\label{eqn.cov.W}
			\DW{\lambda} V^{\mu} &= \pd{\lambda} V^\mu + V^\rho \w{\mu}{\lambda\rho}, 
			\\[2mm] 
			\DW{\lambda} V_{\mu} &= \pd{\lambda} V_\mu - V_\rho \w{\rho}{\lambda\mu}
		\end{align}
	\end{subequations}
	is the covariant derivative of the \We\ connection, and 
	keeping in mind that $ \tetr{a}{\nu}\Laghodge_{\tetr{a}{\mu}} $ is a tensor density of weight $ 
	+1 
	$, balance law \eqref{eqn.EM2} can be rewritten as a covariant divergence with a production term
	\begin{equation*}\label{eqn.EM.cov}
		\DW{\mu}\EMmat{\mu}{\nu} = -\EMmat{\mu}{\nu} \Tors{\rho}{\mu\rho},
	\end{equation*}
	and hence, $ \EMmat{\mu}{\nu} $ is also not conserved in the covariant sense.
	

	For later needs, the following expression of the energy momentum $ \EMmat{\mu}{\nu} =
	\tetr{a}{\nu} \Laghodge_{\tetr{a}{\mu}} $ is required
	\begin{equation}\label{eqn.TEGR.vacuum}
		\EMmat{\mu}{\nu} 
		= 
		2 \HDT{a\lambda\mu}\Laghodge_{\HDT{a\lambda\nu}} - 
		(\HDT{a\lambda\rho}\Laghodge_{\HDT{a\lambda\rho}} - \Laghodge ) 
		\delta^\mu_{\ \nu}.
	\end{equation}
	which is valid for the TEGR Lagrangian discussed in Sec.\,\ref{sec.closure}. 
	This formula will be used later in the 3+1-split and is analogous to
	\cite[Eq.(10.13)]{AldrovandiPereiraBook}.

	\section{Preliminaries for the 3+1 split}\label{sec.31.prep}

	\subsection{Transformation of the torsion equations}\label{sec.transform.potential}

	Before performing a 3+1-split \cite{Alcubierre2008} of system \eqref{eqn.1st.order.TEGR}, 
	we 
	need to do some preliminary transformations of every equation in \eqref{eqn.1st.order.TEGR}.

	Similar to electromagnetism, we introduce the ``electric'' and ``magnetic''
	fields:
	\begin{equation}
		\ET{a}{\mu} := \Tors{a}{\mu\nu} u^\nu, \qquad  \BT{a}{\mu} := \HDT{a\mu\nu} u_\nu
	\end{equation}
	Note that $ \ET{a}{\mu} $ is a tensor, while $ \BT{a}{\mu}
	$ is a tensor-density.
	
	It is known that for any skew-symmetric tensor, its Hodge dual, and a time-like vector $ u^\mu 
	$ 
	the following 
	decompositions hold
	\begin{subequations}\label{eqn.T.decompos}
		\begin{align}
			\HDT{a\mu\nu} &= u^\mu \BT{a}{\nu} - u^\nu \BT{a}{\mu} + 
			\LCsymb^{\mu\nu\lambda\rho}u_\lambda 
			\ET{a}{\rho},\\[2mm]
			\Tors{a}{\mu\nu} &= u_\mu \ET{a}{\nu} - u_\nu \ET{a}{\mu} - 
			\LCsymb_{\mu\nu\lambda\rho}u^\lambda 
			\BT{a}{\rho}.
		\end{align}
	\end{subequations}
	
	Furthermore, we assume that the Lagrangian density can be \textit{equivalently} expressed in 
	different sets of variables, i.e. 
	\begin{equation}\label{eqn.Lagrangians2}
		\Laghodge(\tetr{a}{\mu},\HDT{a\mu\nu}) = \Lagtors(\tetr{a}{\mu},\Tors{a}{\mu\nu}) = 
		\LagBE(\tetr{a}{\mu},\BT{a}{\mu},\ET{a}{\nu}).
	\end{equation}
	It then can be shown that the derivatives of these different representations of the Lagrangian 
	are 
	related as
	\begin{subequations}
		\begin{align}
			\Laghodge_{\HDT{a\mu\nu}}u^\nu &= -\frac12\left( \LagBE_{\BT{a}{\mu}} +u_\mu 
			\LagBE_{\BT{a}{\lambda}} u^\lambda \right), \label{eqn.1st.order.EL}
			\\[2mm]
			\Lagtors_{\Tors{a}{\mu\nu}}u_\nu &= -\frac12\left( \LagBE_{\ET{a}{\mu}} + u^\mu 
			\LagBE_{\ET{a}{\lambda}} u_\lambda \right),
		\end{align}
	\end{subequations}
	and hence, \eqref{eqn.TEGR0.EL} and \eqref{eqn.TEGR0.integr} 
	can be written as (see Appendix~\eqref{app.sec.Deqn})
	\begin{subequations}\label{eqn.tors.BE}
		\begin{align}
			\D{\nu}( u^\mu\LagBE_{\ET{a}{\nu}} - u^\nu \LagBE_{\ET{a}{\mu}} + 
			\LCsymb^{\mu\nu\lambda\rho}u_\lambda\LagBE_{\BT{a}{\rho}}) 
			&= \NC{a}{\mu}\label{eqn.tors.BE.a} \\[2mm]
			\D{\nu}(u^\mu \BT{a}{\nu} - u^\nu\BT{a}{\mu} + 
			\LCsymb^{\mu\nu\lambda\rho}u_\lambda\ET{a}{\rho}) &= 0,
		\end{align}
	\end{subequations}
	where the source $ \NC{a}{\mu} = \Laghodge_{\tetr{a}{\mu}} $ has yet to be developed.
	
	Let us now introduce a new potential $ \Um(\tetr{a}{\mu},\Bm{a}{\mu},\Dm{a}{\mu}) $ as a 
	partial 
	Legendre transform of the Lagrangian $ \LagBE $
	\begin{equation}\label{eqn.Legendre1}
		\Um(\tetr{a}{\mu},\Bm{a}{\mu},\Dm{a}{\mu}) := \ET{a}{\lambda}\LagBE_{\ET{a}{\lambda}} - 
		\LagBE.
	\end{equation}
	By abusing a little bit notations for $ \BT{a}{\mu} $ (we shall use the same letter for $ 
	\BT{a}{\mu} $ and $ -\Bm{a}{\mu} $, this is an intermediate change 
	of variables and will not appear in the final formulation), we introduce the new state variables
	\begin{equation}\label{eqn.Legendre2}
		\Dm{a}{\mu} := \LagBE_{\ET{a}{\mu}}, \qquad \Bm{a}{\mu} := -\BT{a}{\mu}, \qquad 
		\tetr{a}{\mu} := \tetr{a}{\mu}.
	\end{equation}
	Note 
	that both $ \Dm{a}{\mu} $ and $ \Bm{a}{\mu} $ are 
	\emph{tensor-densities}. For derivatives of the new potential, we have the following relations
	\begin{equation}\label{eqn.Legendre3}
		\Um_{\Dm{a}{\mu}} = \ET{a}{\mu}, \quad \Um_{\Bm{a}{\mu}} = \LagBE_{\BT{a}{\mu}},
		\quad \Um_{\tetr{a}{\mu}} = - \LagBE_{\tetr{a}{\mu}}.
	\end{equation}
	This allows us to rewrite equations \eqref{eqn.tors.BE} in 
	the form similar to the non-linear 
	electrodynamics of moving media~\cite{Obukhov2008,DPRZ2017,Hohmann2018a}
	\begin{subequations}
		\begin{align}
			\D{\nu}(u^\mu\Dm{a}{\nu} - u^\nu \Dm{a}{\mu} + 
			\LCsymb^{\mu\nu\lambda\rho}u_\lambda 
			\Um_{\Bm{a}{\rho}})
			& =	\NC{a}{\mu},\\[2mm]
			\D{\nu}(u^\mu \Bm{a}{\nu} - u^\nu \Bm{a}{\mu} - 
			\LCsymb^{\mu\nu\lambda\rho}u_\lambda 
			\Um_{\Dm{a}{\rho}}) 
			& = 0,
		\end{align}
	\end{subequations}
	with $\Bm{a}{\mu}$ and $\Dm{a}{\mu}$ being the analogs of the magnetic induction and
	electric displacement fields, accordingly.

	Finally, we need to express also the Noether current $ \NC{a}{\mu} = \Laghodge_{\tetr{a}{\mu}} 
	$ in 
	terms of the 
	new potential $ \Um $ and the fields $ \Dm{a}{\mu} $ and $ \Bm{a}{\mu} $. One has (see details 
	in Appendix~\ref{app.sec.NC})
	\begin{subequations}\label{eqn.JiA}
		\begin{multline}\label{eqn.Ji0}
			\NC{\indalg{0}}{\mu} = 
				-\Um_{\tetr{\indalg{0}}{\mu}}
				+ u^\lambda \Bm{b}{\mu} \Um_{\Bm{b}{\lambda}} 
				- u^\mu \Bm{b}{\lambda} \Um_{\Bm{b}{\lambda}}\\ 
				- u^\mu \Dm{b}{\lambda} \Um_{\Dm{b}{\lambda}}
				+ \LCsymb^{\mu\lambda\rho\sigma} u_\rho \Um_{\Bm{b}{\lambda}}
				\Um_{\Dm{b}{\sigma}},
		\end{multline}
		\begin{multline}
				\NC{\sA}{\mu} = -\Um_{\tetr{\sA}{\mu}}	\\
				- \itetr{\nu}{\sA}u^\mu
				\left(
				u_\lambda \Dm{b}{\lambda} \Um_{\Dm{b}{\nu}} 
				- \LCsymb_{\nu\lambda\rho\sigma}u^\rho\Bm{b}{\sigma}\Dm{b}{\lambda}
				\right).
		\end{multline}
	\end{subequations}

	\subsection{Transformation of the tetrad equations}
	
	%
	%

	Contracting \eqref{eqn.TEGR0.tetrad} with the 4-velocity $ u^\mu $, and then after 
	change of variables \eqref{eqn.Legendre2} and \eqref{eqn.Legendre3}, the 
	resulting equation reads as
	\begin{equation}
		(\D{\mu}\tetr{a}{\nu} - \D{\nu}\tetr{a}{\mu}) u^\nu = \Um_{\Dm{a}{\mu}}.
	\end{equation}
	Furthermore, using the identity $ \itetr{\mu}{b}\D{\nu}\tetr{a}{\nu} = - 
	\tetr{a}{\nu}\D{\nu}\itetr{\mu}{b}$ and the definition $ u^\mu = \itetr{\mu}{\indalg{0}} 
	$, the 
	latter equation can be rewritten as
	\begin{equation}
		u^\nu\D{\nu}\tetr{a}{\mu} + \tetr{a}{\nu}\D{\mu}u^\nu =-\Um_{\Dm{a}{\mu}},
	\end{equation}
	that later will be used in the 3+1-split.
	
	\subsection{Transformation of the energy-momentum}
	Finally, we express the gravitational part of the energy-momentum tensor $ {\EMmat{\mu}{\nu}} $ 
	\eqref{eqn.TEGR.vacuum} 
	in 
	terms of 
	new 
	variables 
	\eqref{eqn.Legendre2} and 
	the potential $ \Um(\tetr{a}{\mu},\Bm{a}{\mu},\Dm{a}{\mu}) $, while we keep energy-momentum 
	equation \eqref{eqn.EM2} unchanged. It reads
	\begin{align}\label{eqn.sigma.BD}
		{\EMmat{\mu}{\nu}} =
		& -\Bm{a}{\mu}\Um_{\Bm{a}{\nu}} - \Dm{a}{\mu}\Um_{\Dm{a}{\nu}} \nonumber\\
		& - u^\lambda u_\nu \Bm{a}{\mu} \Um_{\Bm{a}{\lambda}} 
		- u^\mu u_\lambda \Dm{a}{\lambda} \Um_{\Dm{a}{\nu}}
		\nonumber\\
		& + u^\mu u_\nu \Bm{a}{\lambda} \Um_{\Bm{a}{\lambda}} 
		+ u^\mu u_\nu \Dm{a}{\lambda} \Um_{\Dm{a}{\lambda}}
		\nonumber \\
		& + \LCsymb_{\nu\sigma\lambda\rho} u^\mu u^\sigma \Bm{a}{\lambda} \Dm{a}{\rho} 
		+ \LCsymb^{\mu\sigma\lambda\rho} u_\nu u_\sigma \Um_{\Bm{a}{\lambda}} 
		\Um_{\Dm{a}{\rho}} 
		\nonumber \\
		& + (\Bm{a}{\lambda} \Um_{\Bm{a}{\lambda}} + \Dm{a}{\lambda} \Um_{\Dm{a}{\lambda}} - 
		\Um) \KD{\mu}{\nu}. 
	\end{align}
	We shall need this expression for $ \EMmat{\mu}{\nu} $ in the last part of the derivation of 
	the $ 
	3+1 $ equations.
	
	
	Therefore, the TEGR system in its intermediate form for the 
	unknowns $ \{\tetr{a}{\mu},\Dm{a}{\mu},\Bm{a}{\mu}\} $  reads
	\begin{subequations}\label{eqn.PDE.4D}
		\begin{align}
			\D{\nu}(u^\mu\Dm{a}{\nu} - u^\nu \Dm{a}{\mu} + 
			\LCsymb^{\mu\nu\lambda\rho}u_\lambda 
			\Um_{\Bm{a}{\rho}})
			& =	\NC{a}{\mu},\label{eT}\\[2mm]
			\D{\nu}(u^\mu \Bm{a}{\nu} - u^\nu \Bm{a}{\mu} - 
			\LCsymb^{\mu\nu\lambda\rho}u_\lambda 
			\Um_{\Dm{a}{\rho}}) 
			& = 0,\label{bT}\\[2mm]
			%
			\pd{\mu}\NC{a}{\mu}
			& = 0, \\[2mm] 
			u^\nu\D{\nu}\tetr{a}{\mu} + \tetr{a}{\nu}\D{\mu}u^\nu &=-\Um_{\Dm{a}{\mu}},
			\label{tetr}
		\end{align}
	\end{subequations}
	with $ \NC{a}{\mu} $ given by \eqref{eqn.JiA}. After we introduce a particular choice of the 
	observer's 4-velocity $ u^\mu $ at the beginning of the next section, we shall finalize 
	transformation of 
	system \eqref{eqn.PDE.4D} to 
	present the final 3+1 equations of TEGR.


	\section{3+1 split of the TEGR equations}	\label{sec.31}

	In this section, we derive a 3+1 version of system \eqref{eqn.PDE.4D} that can be used in a 
	computational code for numerical relativity.
	
	We first recall that Latin indices from the middle of the alphabet $ i,j,k,\ldots=1,2,3 $ are 
	used 
	to 
	denote the spatial components of the 
	spacetime 
	tensors, and Latin indices $ \sA,\sB,\sC,\ldots = \indalg{1},\indalg{2},\indalg{3} $ to denote 
	the 
	spatial 
	directions in the tangent Minkowski space. Additionally, we use the hat on top of a number, 
	e.g. $ 
	\indalg{0} $, for the indices   marking the time and space direction in the tangent space in 
	order 
	to distinguish them 
	from the indices of the spacetime tensors.  Also recall that observer's 4-velocity 
	$ 
	u^\mu $ is associated with the $ \hat{0} $-th column of the 
	inverse tetrad $ \itetr{\mu}{a} $, while the covariant components $ u_\mu $ of the 4-velocity 
	with 
	the $ 
	\hat{0} $-th row of the frame field. For $ u^\mu $ and $ u_\mu $ we standardly assume 
	\cite{Alcubierre2008,RezzollaZanottiBook}:
	\begin{subequations}\label{eqn.4vel}
		\begin{align}
			u^{\mu} &= \itetr{\mu}{\indalg{0}}  = \lapse^{-1}(1,-\shift{i}), 
			\\
			u_\mu   &= - \tetr{\indalg{0}}{\mu} = (-\lapse,0,0,0),
		\end{align}
	\end{subequations}
	with $ \lapse $ being the \emph{lapse function}, and $ \shift{i} $ being the \emph{shift 
	vector}. 
	One can write down $ \tetr{a}{\mu} $ and $ \itetr{\mu}{a} $ explicitly :
	\begin{subequations}\label{eqn.h.eta.matrix}
		\begin{equation}
			\bm{\tetrsymbol} = \left(
			\begin{array}{cccc}
				\alpha          & 0 & 0 & 0 \\[1mm]
				\beta^{\indalg{1}} &  &  &  \\
				\beta^{\indalg{2}} &  & \tetr{\sA}{i} &  \\
				\beta^{\indalg{3}} &  &  & 
			\end{array}
			\right) ,
		\end{equation}
		\begin{equation}
			\bm{h}^{-1} = \bm{\itetrsymbol} = \left(
			\begin{array}{rlcl}
				1/\alpha          & 0 & 0 & 0 \\[1mm]
				-\beta^{1}/\alpha &  &  &  \\
				-\beta^{2}/\alpha &  & \left (\tetr{\sA}{i}\right )^{-1} &  \\
				-\beta^{3}/\alpha &  &  & 
			\end{array}
			\right) ,
		\end{equation}
	\end{subequations}
	where $ \beta^{\sA} = \tetr{\sA}{i}\beta^i$.
	The metric tensor and its inverse are (e.g. see \cite{Gourgoulhon2012a})
	\begin{subequations}
		\begin{align}
			g_{\mu\nu} &= \left(
			\begin{array}{cc}
				-\alpha^2 + \beta_i\beta^i & \beta_i \\[1mm]
				\beta_i & \gamma_{ij}  \\
			\end{array}
			\right) ,
			\\[2mm]
			g^{\mu\nu} &= \left(
			\begin{array}{cc}
				-1/\alpha^2       & \beta^i/\alpha^2 \\[1mm]
				\beta^{i}/\alpha^2 & \gamma^{ij} - \beta^i\beta^j/\alpha^2  \\
			\end{array}
			\right) ,
		\end{align}
	\end{subequations}
	where $ \gamma_{ij} = \kappa_{\sA\sB} \tetr{\sA}{i}\tetr{\sB}{j}$, $ 
	\gamma^{ij} = \left( \gamma_{ij} \right)^{-1}  $, and $ \beta_i =  \gamma_{ij}\beta^j$.
	
	In the rest of the paper, $ \detTetr_3 $ stands for $ \det(\tetr{\sA}{i}) $ so that
	\begin{equation}\label{eqn.det}
		\detTetr := \det(\tetr{a}{\mu}) = \alpha \detTetr_3,
	\end{equation}
	and we use the following convention for the three-dimensional Levi-Civita symbol
	\begin{equation}\label{eqn.LC.3d}
		\LCsymb^{0ijk} = \LCsymb^{ijk}, 
		\qquad
		\LCsymb_{0ijk} =-\LCsymb_{ijk}.
	\end{equation}


	%
	
	\subsection{3+1 split of the torsion PDEs}	\label{ssec.31.tors}
	
	
	\paragraph{Case $ \mu = i=1,2,3 $.} 
	
	For our choice of observer's velocity \eqref{eqn.4vel}, equations \eqref{eT} 
	and \eqref{bT} read
	\begin{subequations}\label{eqn.tpo.1}
		\begin{multline}
			\pd{t} (\lapse^{-1}\Dm{a}{i}) + \pd{k}(\shift{i} 
			\lapse^{-1}\Dm{a}{k} - \shift{k}\lapse^{-1}\Dm{a}{i} \\
			- \LCsymb^{ikj} \lapse \,
			\Um_{\Bm{a}{j}}) 
			= -\NC{a}{i},
		\end{multline}
		\begin{multline}
			-\pd{t} (\lapse^{-1}\Bm{a}{i}) + \pd{k}(\shift{k} 
			\lapse^{-1}\Bm{a}{i} - \shift{i}\lapse^{-1}\Bm{a}{k}  \\
			+ \LCsymb^{ikj} \lapse \,
			\Um_{\Dm{a}{j}}) 
			= 0 .
		\end{multline}
	\end{subequations}
	Because of the factor $ \alpha^{-1} $ everywhere in these equations, it is convenient to 
	rescale 
	the 
	variables:
	\begin{equation}\label{eqn.varDB}
		\aD{a}{\mu} := \lapse^{-1} \Dm{a}{\mu}, \qquad \aB{a}{\mu} := -\lapse^{-1}\Bm{a}{\mu}
	\end{equation}
	so that the derivatives of the potential $ \aU(\tetr{a}{\mu},\aD{a}{\mu},\aB{a}{\mu}) := 
	\Um(\tetr{a}{\mu},\Dm{a}{\mu},\Bm{a}{\mu})
	$ 
	transform as
	\begin{equation}\label{eqn.change.alphaU}
		\aU_{\aD{a}{\mu}} =  \lapse \, \Um_{\Dm{a}{\mu}},
		\qquad
		\aU_{\aB{a}{\mu}} = -\lapse \, \Um_{\Bm{a}{\mu}}.
	\end{equation}
	Hence, \eqref{eqn.tpo.1} can be rewritten as
	\begin{subequations}\label{eqn.tpo.2}
		\begin{align}
			\pd{t} \aD{a}{i} + \pd{k}(\shift{i} 
			\aD{a}{k} - \shift{k}\aD{a}{i}  - \LCsymb^{ikj} \,
			\aU_{\aB{a}{j}}) & 
			= -\NC{a}{i},\\[2mm]
			\pd{t} \aB{a}{i} + \pd{k}(\,\shift{i} 
			\aB{a}{k} - \shift{k}\aB{a}{i}  + \LCsymb^{ikj} 
			\aU_{\aD{a}{j}}) & 
			= 0.
		\end{align}
	\end{subequations}
	
	Finally, extending formally the shift vector as $ \bm{\beta}=(-1,\beta^{k}) 
	$ and introducing 
	the 
	change of variables
	\begin{equation}\label{eqn.varDB.final}
		\Dfin{a}{\mu} := \aD{a}{\mu} + \beta^{\mu} \aD{a}{0}, \qquad \Bfin{a}{\mu} := \aB{a}{\mu}
	\end{equation}
	which are essentially $ 3 $-by-$ 3 $ matrices ($ \Dfin{a}{0} = 0$, $ \Bfin{a}{0} = 
	\Bfin{\indalg{0}}{\mu} = 0 $, and $ \Dfin{\indalg{0}}{k} $ are given by \eqref{eqn.D0.tmp3}),
	we arrive at the final form of the 3+1 equations for the fields $ \Dfin{\sA}{i} $ and $ 
	\Bfin{\sA}{i} $
	\begin{subequations}\label{eqn.PDE.BD.final}
		\begin{align}
			\pd{t} \Dfin{\sA}{i} + \pd{k}(\shift{i} 
			\Dfin{\sA}{k} - \shift{k}\Dfin{\sA}{i}  - \LCsymb^{ikj} \,
			\Hfin{\sA}{j}) & 
			= -\NC{\sA}{i}, \label{eqn.tpo.D}\\[2mm]
			\pd{t} \Bfin{\sA}{i} + \pd{k}(\,\shift{i} 
			\Bfin{\sA}{k} - \shift{k}\Bfin{\sA}{i}  + \LCsymb^{ikj} 
			\Efin{\sA}{j}) & 
			= 0\label{eqn.tpo.B} .
		\end{align}
	\end{subequations}
	where
	\begin{subequations}
		\begin{align}\label{eqn.HE.fin}
			&\Efin{\sA}{j} = \aU_{\aD{\sA}{j}} = \alpha \Um_{\Dm{\sA}{j}}= \alpha \ET{\sA}{j},
			\\
			&\Hfin{\sA}{j} = \aU_{\aB{\sA}{j}} =-\alpha \Um_{\Bm{\sA}{j}}.
		\end{align}
	\end{subequations}
	In terms of $ \{\Bfin{\sA}{k},\Dfin{\sA}{k}\} $ the potential $  
	\aU(\tetr{a}{\mu},\aD{a}{\mu},\aB{a}{\mu}) $ will be denoted as 
	$
	\Ufin(\tetr{a}{\mu},\Bfin{a}{\mu},\Dfin{a}{\mu})  
	$ and, in the case of TEGR, it reads (see \eqref{eqn.Ufin})
	\begin{align}\label{eqn.Ufin0}
		\Ufin(\tetr{\sA}{k},\Bfin{\sA}{k},\Dfin{\sA}{k}) & =  \nonumber\\
		 -\frac{\alpha}{2\detTetr_3} \bigg( 
		& \varkappa \left( \Dfin{\sB}{\sA} \Dfin{\sA}{\sB} - \frac{1}{2} 
		(\Dfin{\sA}{\sA})^2\right) \nonumber\\
		 +
		&\frac{1}{\varkappa} \left( \Bfinmix{\sA}{\sB} \Bfinmix{\sB}{\sA} - \frac{1}{2} 
		(\Bfinmix{\sA}{\sA})^2
		\right)
		\bigg),
	\end{align}
	Alternatively, it can be computed as
	\begin{equation}\label{eqn.U.bdeh}
		\Ufin = \frac{1}{2} ( \Dfin{a}{k} \Efin{a}{k} + \Bfin{a}{k} \Hfin{a}{k} ).
	\end{equation}

	\paragraph{Case $ \mu = 0 $.} 
	
	The $ 0 $-th equations \eqref{eT} and \eqref{bT} are actually not time-evolution equations but  
	pure spatial (stationary) constraints
	\begin{equation}\label{eqn.div.constr}
		\pd{k} \Dfin{a}{k} = \NC{a}{0}, 
		\qquad
		\pd{k} \Bfin{a}{k} = 0.
	\end{equation}
	Their possible violation during the numerical integration of the time-evolution equations 
	\eqref{eqn.PDE.BD.final} is a well-known problem in the numerical analysis of hyperbolic 
	equations 
	with involution constraints. Different strategies to 
	preserve such stationary constraints   
	are known, e.g. constraint-cleaning approach
	\cite{Munz2000,Dedneretal,Dumbser2019} or constraint-preserving integrators
	\cite{Olivares2022,SIGPR2021,oliynyk2025}.

	\subsection{Tetrad PDE}
	
	Because of our choice of the tetrad \eqref{eqn.h.eta.matrix}, we are only interested in the  
	evolution equations for the components $ \tetr{\sA}{k} $. Thus, using the definition of 
	the 4-velocity \eqref{eqn.4vel}, equation \eqref{tetr} can be written as
	\begin{equation}\label{eqn.tetr.3+1}
		\pd{t} \tetr{\sA}{k} - \beta^i \pd{i} \tetr{\sA}{k} - \tetr{\sA}{i} \pd{k} \beta^i 
		= 
		-\Efin{\sA}{k}.
	\end{equation}
	
	This equation can be also written in a slightly different form. After adding $ 0\equiv - 
	\beta^i 
	\pd{k}\tetr{\sA}{i} + \beta^i \pd{k}\tetr{\sA}{i} $ to the left hand-side of 
	\eqref{eqn.tetr.3+1}, 
	one 
	has  
	\begin{equation}\label{eqn.tetr.3+1.2}
		\pd{t} \tetr{\sA}{k} - \pd{k} (\beta^i \tetr{\sA}{i}) - \beta^i (\pd{i}\tetr{\sA}{k} - 
		\pd{k}\tetr{\sA}{i})
		= 
		-\Efin{\sA}{k}.
	\end{equation}
	Finally, using the definition of $ \BT{\sA}{\mu} $ and $ u_\mu $, we have that 
	\begin{multline}
		\BT{\sA}{\mu} = \HDT{\sA\mu\nu} u_\nu = u_\nu \LCsymb^{\mu\nu\alpha\beta} 
		\pd{\alpha}\tetr{\sA}{\beta} = \\
		-
		u_\nu \LCsymb^{\nu\mu\alpha\beta} \pd{\alpha}\tetr{\sA}{\beta} =
		\alpha \LCsymb^{0\mu\alpha\beta} \pd{\alpha}\tetr{\sA}{\beta} 
	\end{multline}
	and hence (use that $ \LCsymb^{0ijk} =\LCsymb^{ijk} $)
	\begin{equation}\label{eqn.B.curlh}
		\Bfin{\sA}{k} = \LCsymb^{kij}\pd{i}\tetr{\sA}{j},
	\end{equation}
	or
	\begin{equation}
		-\beta^i (\pd{i}\tetr{\sA}{k} - 
		\pd{k}\tetr{\sA}{i})
		=
		\LCsymb_{klj}\beta^l \Bfin{\sA}{j}.
	\end{equation}
	Therefore, \eqref{eqn.tetr.3+1.2} can be also rewritten as 
	\begin{equation}\label{eqn.tetr.3+1.3}
		\pd{t} \tetr{\sA}{k} - \pd{k} (\beta^i \tetr{\sA}{i}) =  
		-\Efin{\sA}{k} - \LCsymb_{klj} \beta^l \Bfin{\sA}{j}.
	\end{equation}
	It is not clear yet which one of these equivalent forms
	\eqref{eqn.tetr.3+1}, \eqref{eqn.tetr.3+1.2}, or \eqref{eqn.tetr.3+1.3} at
	the continuous level is more suited for the numerical discretization.
	However, for a structure-preserving integrator, e.g.
	\cite{SIGPR2021,Olivares2022,oliynyk2025}, which is able  to preserve \eqref{eqn.B.curlh}
	up to the machine precision, all these forms are equivalent.

	\subsection{Energy-momentum PDE}\label{sec.energymomentum}
	
	We now turn to the energy-momentum evolution equation
	\begin{equation}\label{eqn.sigma.31}
		\pd{\mu}\EMmat{\mu}{\nu} 
		= \EMmat{\mu}{\lambda} 
		\w{\lambda}{\mu\nu}
	\end{equation}
	and provide its 3+1 version. Because the Lagrangian density in \eqref{eqn.Lagrangians} 
	represents the sum of the gravitational and matter Lagrangian, the energy-momentum tensor is 
	also 
	assumed to be the sum of the gravity and matter parts: $ \EMmat{\mu}{\nu} = 
	\mat{\EMmat{\mu}{\nu}} 
	+ 
	\gra{\EMmat{\mu}{\nu}} $. However, in what follows, we shall omit the matter part keeping in 
	mind 
	that the energy and momentum equations discussed below are equations for the total 
	(matter+gravity) 
	energy-momentum. 
	
	We first, explicitly split \eqref{eqn.sigma.31} into the three momentum and one energy equation:
	\begin{subequations}\label{eqn.EM.sigma}
		\begin{align}
			\pd{t}\EMmat{0}{i} + \pd{k}\EMmat{k}{i}
			& = \EMmat{\mu}{\lambda} 
			\w{\lambda}{\mu i},\\[2mm]
			\pd{t}\EMmat{0}{0} + \pd{k}\EMmat{k}{0}
			& = \EMmat{\mu}{\lambda} 
			\w{\lambda}{\mu 0}.
		\end{align}
	\end{subequations}
	Applying the change of variables \eqref{eqn.varDB}, \eqref{eqn.change.alphaU}, and 
	\eqref{eqn.varDB.final}, expression \eqref{eqn.sigma.BD} for the energy-momentum reduces to the 
	following expressions
	\begin{itemize}
		\item 
		for the total momentum density, $ \sigma^0_{\ i} = \rho_i $ (the matter part is left 
		unspecified and omitted in this paper):
		\begin{equation}\label{eqn.sigma0i}
			\rho_i :=\LCsymb_{ijl}\Dfin{a}{j}\Bfin{a}{l},
		\end{equation}
		%
		\item
		for the momentum flux, $ \EMmat{k}{i}$:
		\begin{multline}\label{eqn.sigmaki}
			\EMmat{k}{i} = -\Bfin{a}{k} \Hfin{a}{i} - \Dfin{a}{k} \Efin{a}{i}  
			- \LCsymb_{ijl} 
			\beta^k \Dfin{a}{j}\Bfin{a}{l} \\
			+ (\Bfin{a}{j} \Hfin{a}{j} + \Dfin{a}{j} \Efin{a}{j} 
			- \Ufin)\KD{k}{i},
		\end{multline}
		\item
		for the energy density:
		\begin{equation}\label{eqn.sigma00}
			\rho_0 :=\EMmat{0}{0} = \LCsymb_{ijl} 
			\beta^i\Dfin{a}{j}\Bfin{a}{l} - \Ufin = \beta^i 
			\rho_i 
			- \Ufin,
		\end{equation}
		\item
		and for the energy flux, $ \EMmat{k}{0} $:
		\begin{multline}\label{eqn.sigmak0}
			{\EMmat{k}{0}} = -\beta^j( \Bfin{a}{k} \Hfin{a}{j} + \Dfin{a}{k} \Efin{a}{j}) \\
			+
			\beta^k( \Bfin{a}{j} \Hfin{a}{j} + \Dfin{a}{j} \Efin{a}{j}) \\
			-
			\LCsymb_{ijl}\beta^k\beta^i\Dfin{a}{j}\Bfin{a}{l}
			-
			\LCsymb^{kjl} \Efin{a}{j}\Hfin{a}{l}.
		\end{multline}
	\end{itemize}
	It is convenient to split the momentum flux $ \EMmat{k}{i} $ and energy
	flux $ \EMmat{k}{0} $ in \eqref{eqn.EM.sigma} into advective and
	constitutive parts, so that the finale form of the 3+1 equations for the
	total energy-momentum \eqref{eqn.EM.sigma} reads
	\begin{subequations}\label{eqn.EM.PDE}
		\begin{align}
			\pd{t}\rho_i & + \pd{k}(-\beta^k \rho_i + \stress{i}{k}) = 
			\rhs{i},\label{eqn.EM.m}\\[2mm]
			\pd{t}\rho_0 & + \pd{k}(-\beta^k \rho_0 + \beta^i \stress{i}{k} 
			- \LCsymb^{kjl} \Efin{a}{j}\Hfin{a}{l} ) = \rhs{0},\label{eqn.EM.e}
		\end{align}
		where the non-matter (gravity + inertia) part of $ \stress{i}{k} $ is given by
		\begin{gather}\label{eqn.flux.const}
			{\stress{i}{k}} = -\Bfin{a}{k} \Hfin{a}{i} - \Dfin{a}{k} \Efin{a}{i} + (\Bfin{a}{j} 
			\Hfin{a}{j} + 
			\Dfin{a}{j} \Efin{a}{j} 
			- \Ufin)\delta^k_{\ i}.
		\end{gather}
		Drawing parallels between electrodynamics and TEGR, one can note the
		presence of the Poynting vector in two forms in \eqref{eqn.EM.PDE}:
		$\Dfin{}{}\times\Bfin{}{}$ in the momentum density in \eqref{eqn.EM.m} and $\Efin{}{}\times\Hfin{}{}$ in the energy flux in \eqref{eqn.EM.e}.

		The source terms in \eqref{eqn.EM.m} and \eqref{eqn.EM.e} are given by
		\begin{align}\label{eqn.source.rhoi}
			\rhs{i} =&-\rho_k \itetr{k}{\sA} \Efin{\sA}{i} + \rho_j \pd{i}\shift{j}
			+\stress{j}{k} \w{j}{ki}
			,\\[2mm]
			\rhs{0} =&- \left(\pd{t} \ln(\alpha) - \shift{k} \pd{k}\ln(\alpha) \right) \Ufin 
			\nonumber\\
			& + \left( \pd{t} \shift{\sA} - \shift{k} \pd{k}\shift{\sA} \right) \rho_j 
			\itetr{j}{\sA} \nonumber\\
			&- \LCsymb^{kjl} \Efin{a}{j}\Hfin{a}{l}\pd{k}\ln(\alpha) + \stress{j}{k} 
			\itetr{j}{\sA} 
			\pd{k}\shift{\sA}. \label{eqn.f0}
		\end{align}
	\end{subequations}

	\subsubsection{Noether current}
	
	Because the energy-momentum $ \EMmat{k}{i} $ was defined as $ \EMmat{k}{i} = \tetr{a}{i} 
	\NC{a}{k} = \tetr{\sA}{i} \NC{\sA}{k} $, we can use the expression for $ \EMmat{k}{i} $ to have 
	an 
	explicit formula for the Noether current. Thus, we have
	\begin{equation}\label{eqn.J.new}
		\NC{\sA}{k} = \itetr{i}{\sA} \EMmat{k}{i} = \itetr{i}{\sA} (-\shift{k} \rho_i + 
		\stress{i}{k} ) 
		= 
		-\shift{k} \rho_{\sA} + \stress{\sA}{k}.
	\end{equation}
	
	In the same way we can find $ \NC{a}{0} $ necessary in \eqref{eqn.div.constr}: $ \NC{a}{0} = 
	\itetr{\lambda}{a} \EMmat{0}{\lambda} = \itetr{0}{a} \EMmat{0}{0} + \itetr{i}{a} \EMmat{0}{i}$, 
	and 
	hence
	\begin{equation}\label{eqn.Ja0}
		\NC{a}{0} = \left\{
		\begin{array}{rl}
			-\alpha^{-1} \Ufin=\rho_{\indalg{0}},	& a=\indalg{0},  \\[3mm] 
			\itetr{i}{\sA} \rho_i = \rho_{\sA}, & \sA=\indalg{1},\indalg{2},\indalg{3}. \\ 
		\end{array} 
		\right.
	\end{equation}

	\subsubsection{Alternative form of the energy-momentum equations}
	
	In \eqref{eqn.sigma.31}, we deliberately use the energy-momentum tensor with
	both spacetime indices because we would like to use the SHTC and Hamiltonian
	structure \cite{SHTC-GENERIC-CMAT} of these equations for designing
	numerical schemes in the future, e.g. \cite{HTC2022,SPH_SHTC}. However, the
	resulting PDEs \eqref{eqn.EM.PDE} do not have a fully conservative form
	preferable for example when dealing with the shock waves in the matter
	fields. Therefore, one may want to use the true conservation law
	\eqref{eqn.Noether.cons} for the total (matter+gravity) Noether current $
	\NC{a}{\mu} $ instead of \eqref{eqn.EM.PDE}. Thus, in 	notations
	\eqref{eqn.J.new}, \eqref{eqn.Ja0}, four conservation laws $
	\pd{\mu}\NC{a}{\mu} = 0 $ now read
	\begin{subequations}
		\begin{align}
			\pd{t} \rho_\sA + & \pd{k} (-\shift{k} \rho_\sA + \stress{\sA}{k}) = 0,\\[2mm]
			\pd{t} \rho_{\indalg{0}} + & \pd{k} (-\shift{k} \rho_{\indalg{0}} 
			- \alpha^{-1} 
			\LCsymb^{kjl}\Efin{a}{j}\Hfin{a}{l} ) = 0.
		\end{align}
	\end{subequations}

	\subsubsection{Evolution of the space volume}
	
	As in the computational Newtonian mechanics \cite{DPRZ2016,SIGPR2021}, the evolution of the 
	tetrad 
	field at the discrete level has to be performed consistently with the volume/mass conservation 
	law. 
	In 
	TEGR, the equivalent to the volume conservation in the Newtonian mechanics is 
	\begin{equation}\label{eqn.pde.det}
		\pd{\mu}(\detTetr u^\mu) = -\detTetr \ET{\mu}{\mu} 
	\end{equation} 
	which can be obtained after contracting \eqref{tetr} with $ \partial \detTetr/\partial 
	\tetr{a}{\mu} 
	= \detTetr
	\itetr{\mu}{a} $, and where $ \ET{\mu}{\mu} = \itetr{\mu}{a} \ET{a}{\mu}$.

	After using \eqref{eqn.4vel}, \eqref{eqn.h.eta.matrix}, and \eqref{eqn.E}, this balance law 
	becomes
	\begin{equation}\label{eqn.h.PDE}
		\pd{t}\detTetr_3 - \pd{k}(\detTetr_3\shift{k} ) =-\frac{\alpha \varkappa}{2} \Dfin{i}{i},
	\end{equation}
	where $ \detTetr_3 =\det(\tetr{\sA}{k}) = \sqrt{\det(\gamma_{ij})} $.

	\section{Summary of the 3+1 TEGR equations}\label{sec.summary}
	Here, we summarize the 3+1 TEGR equations and give explicit expressions for the 
	constitutive 
	fluxes $ \Efin{\sA}{k} $ and
	$ \Hfin{\sA}{k} $ which then close the specification of the entire system.

	\subsection{Evolution equations}
	
	The system of 3+1 TEGR governing equations reads
	\begin{subequations}\label{eqn.3+1.summary}
		\begin{align}
			\pd{t} \Dfin{\sA}{i} + \pd{k}(\shift{i} 
			\Dfin{\sA}{k} - \shift{k}\Dfin{\sA}{i}  - \LCsymb^{ikj} \,
			\Hfin{\sA}{j}) & 
			= \shift{i} \rho_{\sA} - \stress{\sA}{i},
			\label{eqn.3+1.D}
			\\[2mm]
			\pd{t} \Bfin{\sA}{i} + \pd{k}(\,\shift{i} 
			\Bfin{\sA}{k} - \shift{k}\Bfin{\sA}{i}  + \LCsymb^{ikj} 
			\Efin{\sA}{j}) & 
			= 0,
			\label{eqn.3+1.B}
			\\[2mm]
			%
			\pd{t} \rho_\sA +  \pd{k} (-\shift{k} \rho_\sA + \stress{\sA}{k}) &= 
			0,\label{eqn.3+1.m}\\[2mm]
			\pd{t} \rho_{\indalg{0}} + \pd{k} (-\shift{k} \rho_{\indalg{0}} 
			- \alpha^{-1} 
			\LCsymb^{kjl}\Efin{a}{j}\Hfin{a}{l}) &= 0,\label{eqn.3+1.e}\\[2mm]
			%
			\pd{t} \tetr{\sA}{k} - \beta^i \pd{i} \tetr{\sA}{k} - \tetr{\sA}{i} \pd{k} \beta^i 
			& = 
			-\Efin{\sA}{k},
			\label{eqn.3+1.h}
		\end{align}
	\end{subequations}
	
	with the total (matter+gravity) momentum $ \rho_{\sA} = \itetr{i}{\sA}\rho_i $ and the total 
	energy density $ \rho_{\indalg{0}} = -\alpha^{-1} \Ufin = \NC{\indalg{0}}{0} $ computed from 
	\eqref{eqn.Ja0}, and the gravitational part of the momentum flux $ \stress{i}{\sA} = 
	\itetr{k}{\sA} \stress{i}{k}$  computed from \eqref{eqn.flux.const}.

	In particular, the structure of this system resembles very much the structure of the nonlinear 
	electrodynamics of 
	moving media already solved numerically in \cite{DPRZ2017} for example, as well as the 
	structure of 
	the continuum mechanics
	equations with torsion \cite{Torsion2019}. Moreover,
	despite deep conceptual differences,
	it is identical to 
	the new dGREM formulation of the Einstein equations recently pushed forward in 
	\cite{Olivares2022}.
	However, it is important to mention that the Lagrangian approach adopted here
	in principle permits to generalize the formulation to other $f(\Tscal)$ theories.
	A detailed comparison of 
	these two formulations will be a subject of a future paper.
	
	It is also important to note that the structure of system \eqref{eqn.3+1.summary} remains the 
	same 
	independently of the 
	Lagrangian in use as it can be seen from system \eqref{eqn.PDE.4D} where all the constitutive 
	parts 
	are 
	defined through the derivatives of the potential \eqref{eqn.Legendre1}. If the Lagrangian 
	changes, 
	then only the constitutive functions $ \Efin{a}{\mu} $ and $ \Hfin{a}{\mu} $ have to be 
	recomputed.
	
	\subsection{Constitutive relations}
	
	For the TEGR Lagrangian \eqref{eqn.TEGR.Lagr}, $ \Efin{\sA}{k} $ and $ \Hfin{\sA}{k} $ read
	\begin{subequations}\label{eqn.H.E}
		\begin{multline}
			\Hfin{\sA}{k} := 
			-\frac{\alpha}{\varkappa \detTetr_3} 
			\mg{\sA\sB}\mg{\sC\sD} 
			\left( \tetr{\sD}{k} \tetr{\sB}{j}
			- \frac12 \tetr{\sD}{j} \tetr{\sB}{k} \right) 
			\Bfin{\sC}{j}
			\\
			- 
					\frac{1}{\varkappa \detTetr_3} \mg{\sA,\sB} \LCsymb^{ijl} \gamma_{ik} 
			\tetr{\sB}{j} 
					\Efin{\indalg{0}}{l}
		\end{multline}
		\begin{multline}
			\Efin{\sA}{k} := 
			-\frac{\alpha \varkappa}{\detTetr_3} 
			\left( \tetr{\sA}{j} \tetr{\sB}{k} - \frac12 \tetr{\sA}{k} \tetr{\sB}{j} \right) 
			\Dfin{\sB}{j} 
			\\
			- 
					\frac{\varkappa}{\detTetr_3} \LCsymb^{ijl} 
					\gamma_{ik} \tetr{\sA}{j} \Hfin{\indalg{0}}{l}.
			\label{eqn.E}
		\end{multline}
	\end{subequations}
	Note that these relations can also be written as
	\begin{align}\label{eqn.H.E.split}
		\Hfin{\sA}{k} &= 
		\frac{\pd{}\Ufin\hfill}{\pd{}\Bfin{\sA}{k}}
		{ 
			-
			\frac{1}{\varkappa}\detTetr_3 \LCsymb_{kjl} \itetr{j}{\sA} A^l},
		\\[2mm]
		\Efin{\sA}{k} &= 
		\frac{\pd{}\Ufin\hfill}{\pd{}\Dfin{\sA}{k}}
		{
			-
			\varkappa \detTetr_3 \LCsymb_{kjl} 
			\itetr{j}{\sB} \Omega^l}\MG{\sA\sB},
	\end{align}
	where $ A_k = \Efin{\indalg{0}}{k} $ and $ \Omega_k =\Hfin{\indalg{0}}{k}$, and with the potential $ \Ufin $ given by \eqref{eqn.Ufin0}.

	\subsection{Stationary differential constraints}
	
	System \eqref{eqn.3+1.summary} is supplemented by several differential constraints. Thus, as 
	already was mentioned, the $ 0 $-\textit{th} equations ($ \mu=0 $) of \eqref{eT} are not 
	actually 
	time-evolution equations 
	but 
	reduce to the so-called Hamiltonian and momentum stationary divergence-type constraints that 
	must 
	hold on the 
	solution to \eqref{eqn.3+1.summary} at every time instant:
	\begin{equation}\label{eqn.div.constrD}
		\pd{k} \Dfin{\indalg{0}}{k} = \rho_{\indalg{0}},
		\qquad
		\pd{k} \Dfin{\sA}{k} = \rho_{\sA}, 
	\end{equation}
	with $ \rho_{\indalg{0}} = -\alpha^{-1} \Ufin = \NC{\indalg{0}}{0} $.
	
	Accordingly, the $ 0 $-\textit{th} equation of \eqref{bT} gives the divergence constraint on 
	the $ 
	\Bfin{\sA}{i} $ field
	\begin{equation}\label{eqn.div.constrB}
		\pd{k} \Bfin{\sA}{k} = 0,
	\end{equation}
	
	Finally, from the definition of $ \Bfin{\sA}{i} $, we also have a curl-type constraint on 
	the spatial components of the tetrad field:
	\begin{equation}\label{eqn.constrh}
		\LCsymb^{kij}\pd{i}\tetr{\sA}{j} = \Bfin{\sA}{k} .	
	\end{equation}
	

	\subsection{Algebraic constraints}
	As a consequence of our choice of observer's reference frame \eqref{eqn.4vel}, and the fact 
	that $ \ET{a}{\mu} u^\mu = 0 $ and $ \BT{a}{\mu} u_\mu = 0$, we have 
	the following algebraic constraints 
	\begin{subequations}\label{eqn.constr.alg.EB}
		\begin{equation}
			\Bfin{\indalg{0}}{\mu} = 0, \qquad \Bfin{a}{0} = 0,
		\end{equation}
		\begin{equation}\label{eqn.alg.constrE}
			\Efin{\indalg{0}}{0} = \beta^k \Efin{\indalg{0}}{k}, 
			\qquad 
			\Efin{\indalg{0}}{k} = \pd{k} \alpha,
			\qquad  
			\Efin{\sA}{0} = \beta^j \Efin{\sA}{j},
		\end{equation}
	\end{subequations}
	
	\begin{subequations}
		\begin{gather}\label{eqn.D0.tmp3}
			\Dfin{a}{0} = 0,
			\qquad
			\Dfin{\indalg{0}}{k} =-\frac{1}{\varkappa\detTetr_3} \LCsymb^{kli} 
			\Bfinmix{}{li},
			\qquad
			\Dfin{ik}{} = \Dfin{ki}{},
			\\
			\Hfin{\sA}{0} = \shift{k} \Hfin{\sA}{k},
		\end{gather}
	\end{subequations}
	where $ \Dfin{ik}{} = \gamma_{ij} \Dfin{\sA}{j} \tetr{\sA}{k}$.
	As already noted in \cite{AldrovandiPereiraBook}, the gravitational part of the energy-momentum 
	$ 
	\EMmat{\mu}{\nu} $  is trace-free:
	\begin{equation}
		{\EMmat{\mu}{\mu}} = 0.
	\end{equation}

	\section{Hyperbolicity of the vacuum 3+1 TEGR equations}\label{sec.hyperbolicity}

	Due to the presence of the total (gravity + matter) momentum and energy
	equations \eqref{eqn.3+1.m} and \eqref{eqn.3+1.e}, analysis of hyperbolicity
	of the 3+1 TEGR system \eqref{eqn.3+1.summary} requires consideration of a
	model for matter, which however goes beyond the scope of this paper.
	Nevertheless, we can still analyze hyperbolicity of the vacuum TEGR
	equations \eqref{eqn.3+1.summary} for which the momentum and energy
	equations can be omitted. This would be still an important result since in
	the empty space the field equations must be causal and have well-posed
	initial value problem. 
	
	In empty space, the 3+1 TEGR system reduces to the evolution equations
	\eqref{eqn.3+1.D}, \eqref{eqn.3+1.B}, and \eqref{eqn.3+1.h} for the
	gravitational variables $\{\Dfin{\sA}{i},\Bfin{\sA}{i},\tetr{\sA}{i}\}$,
	subject to the constraints \eqref{eqn.div.constrD} and
	\eqref{eqn.div.constrB} and appropriate gauge conditions on
	$\Efin{\indalg{0}}{i}$ and $\Hfin{\indalg{0}}{i}$. In fact, as we shall see,
	the choice of the gauge conditions for $\Efin{\indalg{0}}{i}$ and
	$\Hfin{\indalg{0}}{i}$ plays an important role in our hyperbolicity
	analysis. Moreover, we shall assume that the lapse $\lapse$ and shift
	$\shift{i}$ are not dynamical but some prescribed functions of space and
	time.

	We will show how the vacuum 3+1 TEGR equations, when restricted to
	solutions satisfying the Hamiltonian constraint \eqref{eqn.div.constrD}, can
	be transformed into an equivalent first-order system. This system has the
	principal part of the differential operator identical to that of the
	first-order tetrad reformulation of GR developed by Estabrook, Robinson, and
	Wahlquist \cite{Estabrook1997} and by Buchman and Bardeen
	\cite{Estabrook1997}. This specific formulation, which we will refer to as
	the \ERWBB\ formulation, is known to be \textit{symmetric hyperbolic} if
	subjected to a certain type of gauge conditions. This indirectly
	demonstrates the hyperbolicity of the vacuum TEGR equations. However, this
	result should be taken with care, especially in the context of numerical
	relativity since some differential terms will be turned into algebraic terms
	using the Hamiltonian constraint. This means that the differential operators
	of the original 3+1 TEGR equations \eqref{eqn.3+1.D}, \eqref{eqn.3+1.B}, and
	\eqref{eqn.3+1.h} and the new system having the form of the \ERWBB\
	formulation are not exactly equivalent. In other words, in the context of
	numerical relativity, this may require a constraint-compatible
	discretization, e.g. see \cite[Sec.VI.G]{Olivares2022} or
	\cite{oliynyk2025}.

	Moreover, we could not confirm or disprove the equivalence of the TEGR
	equations written in the \ERWBB\ form and the tetrad \ERWBB\ formulation
	itself. Although their differential parts are the same, we were not able to
	get the algebraic parts of the equations to match. Consequently, we do not
	claim an exact equivalence between the 3+1 TEGR equations and the ERWBB
	tetrad formulation of GR. Our hyperbolicity analysis remains valid, however,
	since it only depends on the principal part of the differential operator.

	Let us also remind that solely hyperbolicity is not enough for
	well-posedness of the initial value problem for a general quasi-linear
	system of first-order equations. At least, to the best of our knowledge,
	there is no an existence and uniqueness theorem for such a class of
	equations in multiple dimensions\footnote{ Despite this, for a system
	describing causal propagation of signals, the hyperbolicity is considered
	as a minimum requirement for numerical discretization.}. In contrary, such a
	theorem exists for symmetric hyperbolic systems, e.g. see \cite{Serre2007}.
	
	Let us first introduce the main elements of the \ERWBB\ formulation which
	relies on the Ricci rotation coefficients
	\begin{align}\label{eqn.Ricci.rot}
		\mathcal{R}_{abc} 
		:= & \bm{\itetrsymbol}_a\cdot\nabla_c \bm{\itetrsymbol}_b \nonumber\\
		= & \itetr{\mu}{a} g_{\mu\nu} \itetr{\lambda}{c} \pd{\lambda} \itetr{\nu}{b} 
		+   \itetr{\mu}{a} g_{\mu\nu} \itetr{\lambda}{c} \Gamma^{\nu}_{\ \lambda\rho} 
		\itetr{\rho}{b}	
	\end{align}
	where $ \nabla_c := \itetr{\lambda}{c} \nabla_{\lambda} $ and $ \nabla_{\lambda} $ is the 
	standard 
	covariant derivative of GR of the symmetric Levi-Civita connection, and $ 
	\Gamma^{\mu}_{\ 
		\nu\lambda} $ are the Christoffel symbols of the Levi-Civita connection.
	
	If $ \Tors{a}{bc} = \Tors{a}{\mu\nu}\itetr{\mu}{b}\itetr{\nu}{c} $ and $ \Tors{}{abc} = \mg{ad} 
	\Tors{d}{bc} $ is the torsion coefficients if it is written in the tetrad basis, then the relation between the Ricci 
	rotation coefficients and torsion can be expressed by the formula
	\begin{equation}\label{eqn.Ricci.Tors}
		\mathcal{R}_{abc} = \frac{1}{2} 
		\left(  
			\Tors{}{abc} + \Tors{}{bca} - \Tors{}{cab}
		\right),
	\end{equation}
	which also shows that the Ricci rotation coefficients equal exactly to the so-called contortion 
	tensor with the opposite sign $ \mathcal{R}_{abc} =-K_{abc} $, e.g. see 
	\cite[Eq.(1.63)]{AldrovandiPereiraBook}.
	
	Then, 24 independent entries of $ \mathcal{R}_{abc} $ are organized into the following state 
	variables
	\begin{equation}\label{eqn.KN}
		\Kbuch{\sC}{\sA} := \mathcal{R}_{\sA\indalg{0}\sC},
		\qquad
		\Nbuch{\sB}{\sA} := \frac{1}{2} \LCtens^{\sA\sC\sD}\mathcal{R}_{\sC\sD\sB}
	\end{equation}
	and 
	\begin{equation}\label{eqn.aw}
		a_{\sA} = \mathcal{R}_{\sA\indalg{0}\indalg{0}},
		\qquad
		\omega^{\sA} = \frac{1}{2} \LCtens^{\sA\sB\sC} \mathcal{R}_{\sC\sB\indalg{0}}
	\end{equation}
	Here, $ \LCtens_{\sA\sB\sC} = 
	\detTetr_3 \LCsymb_{ijk}\itetr{i}{\sA}\itetr{j}{\sB}\itetr{k}{\sC} $.

	Additionally, introducing the vector
	\begin{equation}
		n^\sA = \frac{1}{2} \LCtens^{\sA\sB\sC} \mg{\sC\sD} \Nbuch{\sB}{\sD},
	\end{equation}
	the Hamiltonian constraint in terms of \ERWBB\ takes the form (see \cite[eq.(A1)]{Buchman2003})
	\begin{multline}\label{eqn.Hamiltonian.ERWBB}
		\pd{\sA} n^\sA = \frac12 \Nbuch{\sA\sB}{}\Nbuchmix{\sA\sB}{} + \frac14\left( \Kbuchmix{\sA\sB}{}\Kbuch{\sA}{\sB} - \Nbuch{\sA\sB}{}\Nbuchmix{\sB\sA}{} \right) 
		\\
		- \frac14\left( \left( \Kbuchmix{\sA}{\sA} \right)^2 + \left( \Nbuch{\sA}{\sA} \right)^2\right),
	\end{multline}
	where $\pd{\sA} = \itetr{k}{\sA}\pd{k}$.

	It can be shown that the following one-to-one relations between the \ERWBB\ and TEGR state variables hold
	\begin{subequations}\label{eqn.Buchman.TEGR.KN}
		\begin{align}
			\Kbuch{\sA}{\sB} = & \frac{\varkappa}{\detTetr_3} 
			\left(
			\Dfin{\sA\sB}{}-\frac{1}{2} \Dfin{\sC}{\sC} \mg{\sA\sB}
			\right),\\[2mm]
			-\Nbuch{\sA}{\sB} =&\frac{1}{\detTetr_3}
			\left(
			\Bfinmixx{\sA}{\sB} - \frac{1}{2}\Bfinmix{\sC}{\sC}\KD{\sB}{\sA}
			\right),
		\end{align}
	\end{subequations}
	where $\Dfin{\sA\sB}{} = \mg{\sA\sC}\tetr{\sB}{i}\Dfin{\sB}{i}$ and $\Bfinmixx{\sA}{\sB} = \mg{\sA\sC}\tetr{\sB}{i}\Bfin{\sC}{i}$. We remark that due to our choice of the 3+1 split \eqref{eqn.h.eta.matrix},
	i.e. that observer's time vector, $ \bas{\indalg{0}} $, is aligned with the
	normal vector to the spatial hypersurfaces, the matrices $ \Kbuch{\sA}{\sB}
	$ and $ \Dfin{\sA\sB}{} $ are \textit{symmetric}.
	
	Additionally, we have the following relations between the \ERWBB\ vectors $
	a_{\sA} $, $ \omega^{\sA} $ and the \tegr\ vectors $
	A_{\sA}=\itetr{k}{\sA}\Efin{\indalg{0}}{k} = \pd{k}\lapse $ and $ \Omega_{\sA} =
	\itetr{k}{\sA}\Hfin{\indalg{0}}{k} $
	\begin{equation}\label{eqn.Buchman.TEGR.AW}
		a_{\sA} = \lapse^{-1} A_{\sA},
		\qquad
		\omega^{\sA} =-\varkappa \lapse^{-1} \MG{\sA\sB} \Omega_{\sB}.
	\end{equation}
	
	Finally, expression of the 
	constitutive fluxes $ \Efin{\sA}{k} $ and $ \Hfin{\sA}{k} $ in terms of $ 
	\{\Kbuch{\sA}{\sB},\Nbuchdown{\sA}{\sB},a_{\sA},\omega_{\sA}\} $ read
	\begin{subequations}\label{eqn.const.KN}
		\begin{align}
			\Efin{}{\sA\sB} &=
			- \lapse \,\Kbuch{\sA}{\sB} 
			+ \lapse \LCtens_{\sA\sB\sC} \omega^\sC ,
			\\[2mm]
			\Hfin{\sB}{\sA} &= 
			\phantom{-}\frac{\lapse}{\varkappa} \Nbuchdown{\sA}{\sB} 
			- \frac{\lapse}{\varkappa} 
			\LCtens_{\sA\sB\sC} a^\sC ,
		\end{align}
	\end{subequations}
	where $\Efin{}{\sA\sB} = \mg{\sA\sC}\itetr{k}{\sB}\Efin{\sC}{k}$ and $\Hfin{\sA}{\sB} = \itetr{k}{\sA}\Hfin{k}{\sB}$.
	
	First of all, let us note that, in the tetrad frame, the tetrad equation \eqref{eqn.3+1.h} reads
	\begin{equation}
		\pd{\indalg{0}} \itetr{k}{\sA} = \frac{1}{\lapse} \left( \Efin{\sA}{i}\itetr{i}{\sB}\itetr{k}{\sA} - \pd{\sA} \shift{k} \right),
	\end{equation}
	where $ \pd{\indalg{0}} = \itetr{\mu}{\indalg{0}}\pd{\mu} = u^\mu\pd{\mu} =
	\lapse^{-1}\pd{t} - \lapse^{-1}\shift{k}\pd{k}$, and $ \pd{\sA} =
	\itetr{k}{\sA}\pd{k} $. For a non-dynamical shift $ \shift{k} $, this is
	simply an ordinary differential equation for the frame field $
	\itetr{k}{\sA} $ along the observer's trajectories and hence, it does not
	affect the hyperbolicity analysis.

	Now, rewriting the 3+1 TEGR equations on $\Dfin{\sA}{i}$ and $\Bfin{\sA}{i}$
	(eqs.~\eqref{eqn.3+1.D} and \eqref{eqn.3+1.B}) in terms of $\Dfin{\sA}{\sB}
	= \tetr{\sB}{i}\Dfin{\sA}{i}$ and $\Bfin{\sA}{\sB} =
	\tetr{\sB}{i}\Bfin{\sA}{i}$, then applying transformations
	\eqref{eqn.Buchman.TEGR.KN} to these equations and using relations
	\eqref{eqn.Buchman.TEGR.KN}--\eqref{eqn.const.KN}, after a lengthy but
	rather straightforward sequence of transformations, one obtains the following equations
		\begin{subequations}\label{eqn.KN.pde0}
		\begin{align}
			\frac{1}{\lapse}\pd{\indalg{0}} \Kbuch{\sA}{\sB}
			&
			- \mg{\sA\sC}\LCtens^{\sC\sD\sE} \pd{\sD} \Nbuchdown{\sE}{\sB}
			- \pd{\sA} a_{\sB}  + \mg{\sA\sB} \pd{\sC} n^\sC = \text{l.o.t.}
			\\[2mm]
			\frac{1}{\lapse}\pd{\indalg{0}} \Nbuchdown{\sA}{\sB}
			&
			+ \mg{\sB\sC}\LCtens^{\sC\sD\sE} \pd{\sD} \Kbuch{\sE}{\sA}
			+ \pd{\sA} \omega_{\sB} = \text{l.o.t.}
			\label{eqn.KN.pde.N}
		\end{align}
	\end{subequations}
	where 'l.o.t.' stands for 'low-order terms' (i.e. algebraic terms that do
	not contain space and time derivatives).

	Finally, using the Hamiltonian constraint \eqref{eqn.Hamiltonian.ERWBB}, the
	divergence term $\pd{\sC}n^\sC$ can be transformed into algebraic terms, and
	moved to the right-hand side of the equations. Thus, the 3+1 TEGR equations
	for $\Dfin{\sA}{i}$ and $\Bfin{\sA}{i}$ become
	\begin{subequations}\label{eqn.KN.pde}
		\begin{align}
			\pd{\indalg{0}} \Kbuch{\sA}{\sB}
			&
			- \alpha \mg{\sA\sC}\LCtens^{\sC\sD\sE} \pd{\sD} \Nbuchdown{\sE}{\sB}
			- \alpha \pd{\sA} a_{\sB} = \text{l.o.t.}\label{eqn.KN.pde.K}
			\\[2mm]
			\pd{\indalg{0}} \Nbuchdown{\sA}{\sB}
			&
			+ \alpha \mg{\sB\sC}\LCtens^{\sC\sD\sE} \pd{\sD} \Kbuch{\sE}{\sA}
			+ \alpha \pd{\sA} \omega_{\sB} = \text{l.o.t.}
			\label{eqn.KN.pde.N}
		\end{align}
	\end{subequations}
	We note the opposite order of the subscripts $\sA$ and $\sB$ in
	\eqref{eqn.KN.pde.N} in the second term with respect to
	\cite[eq.(40)]{Buchman2003}. This is solely conditioned by our initial
	choice of ordering the spacetime and tetrad indices in $\Bfin{\sA}{i}$ and
	it does not affect the symmetric hyperbolicity of the system discussed
	below.

	Similar to $ A_k $ and $ \Omega_k $ in TEGR, the vectors $ a_{\sB} $ and $
	\omega_{\sA} $ in the \ERWBB\ formulation of GR are not provided with
	particular evolution equations following from the variational formulation,
	and therefore they are considered as gauge conditions. Hence, one could try
	to choose their evolution equations in such a way as to guaranty the
	well-posedness of the enlarged system for the unknowns $
	\{\Kbuch{\sA}{\sB},\Nbuchdown{\sA}{\sB},a_\sB,\omega_\sB \}$. Thus, as shown
	in \cite{Estabrook1997,Buchman2003}, if coupled with the Nester or Lorentz
	gauge conditions on $a_\sA$ and $\omega_\sA$ having the form:
	\begin{subequations}\label{eqn.aw.pde}
		\begin{align}
			\pd{\indalg{0}} a_\sA - \alpha \MG{\sB\sC}\pd{\sB} \Kbuch{\sC}{\sA} &= \text{l.o.t.},
			\\[2mm]
			\pd{\indalg{0}} \omega_\sA + \alpha \MG{\sB\sC}\pd{\sB} \Nbuchdown{\sC}{\sA} &= \text{l.o.t.},
		\end{align}
	\end{subequations}
	the resulting system \eqref{eqn.KN.pde}--\eqref{eqn.aw.pde} is symmetric
	hyperbolic, i.e. it can be written in a quasi-linear form
	\begin{equation}\label{eqn.quasi.lin}
		\pd{t} \bm{Q} + \bm{M}^k\pd{k} \bm{Q} = \text{l.o.t.}
	\end{equation}
	with matrices $ \bm{M}^k = \bm{M}^k(\alpha,\itetr{i}{\sA})$ being symmetric for arbitrary $\itetr{i}{\sA}$.
	To see this, one needs to order the entries of
	$\Kbuchmix{\sA}{\sB}$ and $\Nbuchdown{\sA}{\sB}$ in the following way
	$\bm{Q} = \{\Kbuch{\indalg{1}}{\sA},\Kbuch{\indalg{2}}{\sA},\Kbuch{\indalg{3}}{\sA},\Nbuchdown{\sA}{\indalg{1}},\Nbuchdown{\sA}{\indalg{2}},\Nbuchdown{\sA}{\indalg{3}},a_\sA,\omega_\sA\}$.
	
	In summary, the 3+1 \tegr\ equations in vacuum, eqs. \eqref{eqn.3+1.D},
	\eqref{eqn.3+1.B}, and \eqref{eqn.3+1.h}, if coupled with proper gauge
	conditions on $A_k$ and $\Omega_k$, are equivalent to symmetric-hyperbolic
	system \eqref{eqn.KN.pde}--\eqref{eqn.aw.pde} on solutions satisfying the
	Hamiltonian constraint. This demonstrates that these equations are causal
	and have well-posed initial value problem under the mentioned conditions.

	Finally, it is worth commenting on the matter coupling. As we will
	discuss in the next section, $\Kbuch{\sA}{\sB}$ is the extrinsic curvature
	of the spatial hypersurfaces. Although the \ERWBB\ formulation was initially
	derived for empty spacetimes \cite{Estabrook1997,Buchman2003}, we can assume
	that the matter stress tensor will appear on the right-hand side of
	\eqref{eqn.KN.pde.K} as a source term, similar to the standard 3+1
	decomposition of the Einstein equations (see
	\cite[eq.(5.19)]{Gourgoulhon2012a}). Similarly, the matter stress tensor is
	present on the right-hand side of the equation for $\Dfin{\sA}{i}$ and will
	therefore be inherited as a source term in \eqref{eqn.KN.pde.K}. This
	maintains the similarity between the two formulations even when matter
	fields are included.
	
	However, if the matter sector is retained, extra equations (such as
	\eqref{eqn.3+1.m} and \eqref{eqn.3+1.e} in case of ideal fluids) must be
	added to the symmetric hyperbolic system defined by
	\eqref{eqn.KN.pde}--\eqref{eqn.aw.pde} in both TEGR and \ERWBB\
	formulations. This addition will likely destroy the symmetric hyperbolicity
	of the enlarged system. Nevertheless, because the relativistic equations for
	ideal fluids are themselves symmetric hyperbolic \cite{Ruggeri1981Euler},
	there is still hope that the combined system for gravity and matter could be
	symmetrized, at least in this simple case.

	\section{Relation between the torsion and extrinsic curvature}
	
	It is useful to relate the state variables of TEGR to the conventional quantities used in 
	numerical 
	general relativity \cite{ADM2008,Baumgarte2003a,Gourgoulhon2012a}. In 
	what follows, we relate the spatial extrinsic curvature of GR to the $ \Dfin{\sA}{k} $ field. 
	We remark 
	that the two fields are conceptually different due to the different geometry interpretations in 
	GR and TEGR. Therefore, the obtained relation is possible only due to the equivalence of TEGR 
	and GR. 
	
	In GR, the evolution equation of the spatial metric $ \gamma_{ij} $ is (see 
	\cite[Eq.(7.64)]{RezzollaZanottiBook})
	\begin{equation}\label{eqn.K}
		\pd{t} \gamma_{ij} - \shift{l}\pd{l}\gamma_{ij} - \gamma_{il}\pd{j}\shift{l}
		-\gamma_{jl}\pd{i}\shift{l} = - 2 \alpha K_{ij},
	\end{equation}
	where  $ K_{ij} $ is the spatial extrinsic 
	curvature.
	On the other hand, in TEGR, contracting \eqref{eqn.tetr.3+1} with $ \mg{\sA\sB} \tetr{\sB}{j} 
	$, one 
	obtains 
	the following evolution equation for the spatial metric:
	\begin{multline}\label{eqn.gamma.PDE}
		\pd{t} \gamma_{ij} - \shift{l}\pd{l}\gamma_{ij} - \gamma_{il}\pd{j}\shift{l}
		-\gamma_{jl}\pd{i}\shift{l} = \\
		-\mg{\sA\sB}(\tetr{\sA}{i} \Efin{\sB}{j} + \tetr{\sA}{j} 
		\Efin{\sB}{i}).
	\end{multline} 
	Therefore, one can deduce an expression for the extrinsic curvature in terms of $ 
	\Efin{\sA}{i} $: 
	\begin{equation}\label{key}
		K_{ij} = \frac{1}{2\alpha}\mg{\sA\sB}(\tetr{\sA}{i} \Efin{\sB}{j} + \tetr{\sA}{j} 
		\Efin{\sB}{i}).
	\end{equation}
	
	To obtain another expression for $ K_{ij} $ in terms of the primary state variable $ 
	\Dfin{\sA}{i} 
	$, one needs to use the constitutive relationship \eqref{eqn.E} to deduce
	\begin{equation}
		K_{ij} = -\frac{\varkappa}{\detTetr_3} 
		\left ( 
		\Dfin{ij}{} - \frac{1}{2} \Dfin{k}{k} \gamma_{ij}
		\right ) ,
	\end{equation}
	which, if contracted, gives the relationship for the traces $ K^i_{\ i} = \gamma^{ij}K_{ji} $ 
	and $ 
	\Dfin{i}{i} =  \Dfin{\sA}{i}\tetr{\sA}{i} $
	\begin{equation}
		K^i_{\ i} = \frac{\varkappa}{2\detTetr_3}\Dfin{i}{i}.
	\end{equation}
	
	Remark that if written in the tetrad frame $ K_{\sA\sB} = 
	\itetr{i}{\sA}\itetr{j}{\sA} K_{ij}$ then the extrinsic curvature is exactly $ \Kbuch{\sA}{\sB} 
	$ 
	introduced in 
	\eqref{eqn.Buchman.TEGR.KN} apart from the opposite sign
	\begin{equation}
		K_{\sA\sB} = -\Kbuch{\sA}{\sB}.
	\end{equation}

	\section{Torsion scalar}\label{sec.closure}

	In \tegr\ and its $ f(\Tscal) $-extensions, the Lagrangian density is a function of the 
	torsion scalar $ \Tscal $, e.g. in \tegr, the Lagrangian density is
	\begin{subequations}\label{eqn.TEGR.Lagr}
		\begin{equation}
			\Lagtors(\tetr{a}{\mu},\Tors{a}{\mu\nu}) = \frac{\detTetr}{2 \, \varkappa} \Tscal,
		\end{equation}
		\begin{align}\label{eqn.tors.scal0}
			\Tscal(\tetr{a}{\mu},\Tors{a}{\mu\nu}) &:= \\
			&\frac14 g^{\beta\lambda} g^{\mu\gamma} g_{\alpha\eta} \Tors{\alpha}{\lambda\gamma}
			\Tors{\eta}{\beta\mu} \\
			&+
			\frac12 g^{\mu\gamma} \Tors{\lambda}{\mu\beta} \Tors{\beta}{\gamma\lambda} \\
			&- 
			g^{\mu\lambda} \Tors{\rho}{\mu\rho} \Tors{\gamma}{\lambda\gamma},	  
		\end{align}
	\end{subequations}
	where $ \varkappa = 8\pi G c^{-4} $ is the Einstein gravitational constant, $ G $ is the 
	gravitational constant, $ c $ is the speed of light in vacuum.


	However, to close system \eqref{eqn.3+1.summary}, we need not the Lagrangian $ 
	\Lagtors(\tetr{a}{\mu},\Tors{a}{\mu\nu})  $ directly but we need to perform a sequence of 
	variable 
	and potential changes: $ \Lagtors(\tetr{a}{\mu},\Tors{a}{\mu\nu}) = 
	\LagBE(\tetr{a}{\mu},\ET{a}{\mu},\BT{a}{\mu}) $ $ \longrightarrow $ $ \ET{a}{\mu} 
	\Dm{a}{\mu} - \LagBE = 
	\Um(\tetr{a}{\mu},\Dm{a}{\mu},\Bm{a}{\mu}) = 
	\Ufin(\tetr{a}{\mu},\Dfin{a}{\mu},\Bfin{a}{\mu}) $. 
	Thus, we have
	\begin{align}\label{eqn.Tscal.EB}
		&\LagBE(\tetr{a}{\mu},\ET{a}{\mu},\BT{a}{\mu}) = \nonumber\\
		&\frac{\detTetr}{2 \, \varkappa}
		\bigg(
		- \frac12(\ETmix{\indalg{0}\lambda}{}\ETmix{\indalg{0}}{\ \lambda} -  
		2\ETmix{\lambda}{\ \,\lambda}\ETmix{\beta}{\ \,\beta} +
		\ETmix{\ \,\lambda}{\beta} \ETmix{\beta}{\ \,\lambda}  +
		\ETmix{\lambda}{\ \,\beta}\ETmix{\beta}{\ \,\lambda} )
		\nonumber
		\\
		& + \LCsymb_{\lambda\gamma\eta\rho} u^\eta 
		(\ETmix{\lambda\gamma}{}\BTmix{\indalg{0}\rho}{} + 2 
		\ET{\indalg{0}\lambda}{}\BTmix{\gamma\rho}{})
		\nonumber
		\\
		&- \frac12 h^{-2} ( \BTmix{\indalg{0}\lambda}{}\BTmix{\indalg{0}}{\ \lambda}
		+ \BTmix{\lambda}{\ \,\lambda}\BTmix{\beta}{\ \,\beta}
		- 2\BTmix{\lambda}{\ \,\beta}\BTmix{\beta}{\ \,\lambda}
		)
		\bigg).
	\end{align}
	%
	
	In turn, if we perform the Legendre transform $ \Um(\tetr{\sA}{k},\Dm{\sA}{k},\Bm{\sA}{k}) = 
	\ET{a}{\mu} 
	\Dm{a}{\mu} - \LagBE$
	then the new potential $ \Um $ reads
	\begin{align}\label{eqn.Lagr.DB}
		\Um(\tetr{\sA}{k},\Dm{\sA}{k},\Bm{\sA}{k}) &= \nonumber\\ 
	  	\frac{1}{4\,h} \bigg( &\varkappa \left(
		\Dm{i}{i}\Dm{k}{k} - 2 \Dm{i}{k}\Dm{k}{i}
		\right)\nonumber\\
		 + &\frac{1}{\varkappa} \left ( 
		\Bmmix{i}{\ \,i}\Bmmix{k}{\ \,k}
		- 2 \Bmmix{i}{\ \,k}\Bmmix{k}{\ \,i}
		\right ) \nonumber\\
		+ \varkappa (\beta^j \Dm{j}{0} + &2 \beta^j \Dm{j}{0}\Dfin{k}{k} - 4 \beta^j \Dm{k}{0} 
		\Dfin{j}{k}) 
		\bigg),
	\end{align}
	where $ \Dm{k}{i} = \Dm{\sA}{i} \tetr{\sA}{k} $ and $ \Bmmix{k}{\ i} = \Bm{\sA}{j} 
	\itetr{k}{\sA} 
	\gamma_{ji} $, and in the last two terms, one should pay attention that the new field $ 
	\Dfin{\sA}{k} $ introduced in \eqref{eqn.varDB.final} appears there.
	
	Apparently, apart from the last terms in \eqref{eqn.Lagr.DB} depending on $ \Dm{k}{0} $, the 
	potential 
	$ \Um $ is more symmetric in the variables $ \Dm{a}{\mu} $ and $ 
	\Bm{a}{\mu} $ rather than the Lagrangian $ \LagBE $ in $ \ET{a}{\mu} $ and $ \BT{a}{\mu} $. 
	This in 
	particular, justifies the introduction of the new and final variables \eqref{eqn.varDB.final}
	\begin{equation}\label{eqn.db}
		\Dfin{\sA}{k} = \Dm{\sA}{k} + \beta^k \Dm{\sA}{0}, \qquad \Bfin{\sA}{k} = \Bm{\sA}{k},
	\end{equation}
	so that the potential $ \Um(\tetr{\sA}{k},\Dm{\sA}{k},\Bm{\sA}{k}) = 
	\Ufin(\tetr{\sA}{k},\Dfin{\sA}{k},\Bfin{\sA}{k})$ becomes just
	\begin{align}\label{eqn.Ufin}
		\Ufin(\tetr{\sA}{k},\Dfin{\sA}{k},\Bfin{\sA}{k}) &= \nonumber\\
		-\frac{\alpha}{2\detTetr_3} \bigg( 
		&\varkappa \left( \Dfin{\sB}{\sA} \Dfin{\sA}{\sB} 
		- \frac{1}{2} (\Dfin{\sA}{\sA})^2\right)\nonumber\\
		+
		&\frac{1}{\varkappa} \left( \Bfinmix{\sA}{\sB} \Bfinmix{\sB}{\sA} - \frac{1}{2} 
		(\Bfinmix{\sA}{\sA})^2
		\right)
		\bigg).
	\end{align}

	\section{Conclusion and discussion}\label{sec.conclusion}
	
	We have presented a 3+1-split of the TEGR equations in their historical pure
	tetrad version, i.e. with the spin connection set to zero. To the best of
	our knowledge, there were not many attempts to obtain a 3+1-split of the
	TEGR equations, e.g. \cite{Maluf2001a,Capozziello2021,Pati2022}, and we are not aware
	of any attempts to solve numerically the full TEGR system of equations for
	general spacetimes. This paper, therefore, may help to cover this gap.
	However, first attempts to solve similar equations for the dGREM equations
	\cite{Olivares2022} in vacuum were done recently \cite{oliynyk2025} with a
	structure compatible discretization.
	
	Our derivation started from the action integral with a
	Lagrangian denisty as an arbitrary function of the tetrad fields and their first gradietns, and therefore some elements of our derivation might be
	used for deriving 3+1 equations of extensions of TEGR such as $ f(\Tscal)
	$-teleparallel theories in the future. After separating the spatial and time
	components of the torsion, the 3+1 governing partial differential equations
	have appeared to have similar structure to the equations of nonlinear
	electrodynamics \cite{DPRZ2017} and equations for continuum fluid and solid
	mechanics with torsion \cite{Torsion2019} if, howeveer, the gauge vectors
	$A_k$ and $\Omega_k$ are set to $0$. For more general gauge conditions on
	$A_k$ and $\Omega_k$, the 3+1 equations have a structure of the Maxwell
	equations coupled with the acoustic-type equations, e.g. see
	\cite{MaxwellGLM}. Moreover, it has appeared that the derived equations are
	equivalent to the recently proposed dGREM tetrad formulation of GR
	\cite{Olivares2022}. However, we emphasize that that the starting point of
	\cite{Olivares2022} was different. Here, we started from the variational
	principle while the fundamental elements of the dGREM formulation are the
	frame field, their exterior derivatives, and the Nester-Witten and Sparling
	forms.
	
	The derived 3+1 TEGR equations are not immediately hyperbolic as usually the
	case for many first-order reductions of the Einstein equations. We
	demonstrated that for the empty space, the differential operator of the 3+1
	vacuum TEGR equations can be transformed into a different but equivalent
	quasi-linear form, provided that the Hamiltonian constraint is fulfilled.
	This form is equivalent to the symmetric hyperbolic tetrad reformulation of
	GR by Estabrook-Robinson-Wahlquist \cite{Estabrook1997} and Buchman-Bardeen
	\cite{Buchman2003} if certain type of gauge conditions is imposed on the
	vectors $\Efin{\indalg{0}}{k}$ and $\Hfin{\indalg{0}}{k}$. The question of
	hyperbolicity of the full 3+1 TEGR equations coupled with matter is still
	open and requires further investigation.
	
	
	Despite it is argued that TEGR is fully equivalent to Einstein's general
	relativity, the proposed 3+1 TEGR equations have yet to be carefully tested
	in a computational code and have yet to be proved to pass all the standard
	benchmark tests of GR. Therefore, further research will concern
	implementation of the TEGR equations in a high-order discontinuous Galerkin
	code \cite{Dumbser2018a,Busto2020}, with a possibility of constraint
	cleaning  \cite{Dumbser2019} and well-balancing \cite{Gaburro2021}. This in
	particular would allow a direct comparison of the TEGR with other 	3+1
	equations of GR, such as Z4 formulations \cite{Alic2012} forwarded by Bona
	\textit{et al} in \cite{Z4}, and FO-CCZ4 by Dumbser \textit{et al}
	\cite{FO-CCZ4}, a strongly hyperbolic formulations of GR, within the same
	computational code.  Another numerical strategy would be to use staggered
	grids and to develop a structure-preserving discretization
	\cite{SIGPR2021,Olivares2022,Fambri2020a,oliynyk2025} that should allow to
	keep errors of div and curl-type involution constraints of TEGR at the
	machine precision. Finally, from the theoretical perspective, it would be
	important to extend the presented technique to covariant teleparallel
	geometries (non-zero spin connection) and $f(\Tscal)$ teleparallel theories
	and compare the 3+1 equations with their Hamiltonian formulations
	\cite{li2011d,li2011e,ferraro2018,blagojevic2020,bajardi2025}.

	\paragraph{Acknowledgments}
	I.P. is grateful to A.~Golovnev for sharing his thoughts about various
	aspects of teleparallel gravity that were important for the appearance of
	this work. Also, I.P. would like to thank M.~Dumbser, O.~Zanotti, E.R. Most
	for their support and insightful discussions. I.P. acknowledges the
	financial support received from the European Union - Next Generation EU,
	Mission 4 Component 2 - CUP E53D23005840006, the Italian Ministry of
	University and Research (MUR) with the PRIN Project 2022 No. 2022N9BM3N.
	During part of the development of this work,
	funding for H.O. came from Radboud University Nijmegen through a Virtual
	Institute of Accretion (VIA) postdoctoral fellowship from the Netherlands
	Research School for Astronomy (NOVA).
	HO is supported by the Individual CEEC program - 5th edition funded by
	the Portuguese Foundation for Science and Technology (FCT). This work is
	supported by the Center for Research and Development in Mathematics and
	Applications (CIDMA) through the Portuguese Foundation for Science and
	Technology (FCT) Multi-Annual Financing Program for R\&D Units (UID/04106).
	The authors acknowledge support from the projects
	CERN/FIS-PAR/0024/2021 and 2022.04560.PTDC. This work has further been
	supported by the  European Horizon Europe staff exchange (SE) programme
	HORIZON-MSCA-2021-SE-01 Grant No. NewFunFiCO-10108625.

	The work of E.R. is supported by the Mathematical Center in Akademgorodok
	under the agreement No. 075-15-2025-348 with the Ministry of Science and
	Higher Education of the Russian Federation.

	\appendix

	\section{Transformation of Noether's current $ \NC{a}{\mu} $}\label{app.sec.NC}
	
	Here, we express Noether's current $ \NC{a}{\mu} = \Laghodge_{\tetr{a}{\mu}} $ for the 
	gravitational part of the Lagrangian (i.e. the matter part is ignored in this section) in terms 
	of 
	the 
	potential $ \Um $ and new variables $ \Dm{a}{\mu} $ and $ \Bm{a}{\mu} $. 
	
	Thus, for the parametrization
	$ \Laghodge(\tetr{a}{\mu},\HDT{a\mu\nu}) = \LagBE(\tetr{a}{\mu},\BT{a}{\mu},\ET{a}{\mu}) $, one 
	has
	\begin{equation}\label{app.eqn.Noether1}
		\Laghodge_{\tetr{a}{\mu}} = \LagBE_{\tetr{a}{\mu}} 
		+ \LagBE_{\BT{b}{\lambda}} \frac{\partial \BT{b}{\lambda}}{\partial \tetr{a}{\mu}}
		+ \LagBE_{\ET{b}{\lambda}} \frac{\partial \ET{b}{\lambda}}{\partial \tetr{a}{\mu}}.
	\end{equation}
	Then, using the definitions of the frame 4-velocity $ u^\nu = \itetr{\nu}{\indalg{0}} $ and $ 
	u_\nu 
	= 
	-\tetr{\indalg{0}}{\nu} $ and the torsion fields
	$ \BT{b}{\lambda} = \HDT{b\lambda\nu} u_\nu = - \HDT{b\lambda\nu} \tetr{\indalg{0}}{\nu}$ and 
	$ \ET{b}{\lambda} = \Tors{b}{\lambda\nu} u^\nu = 
	-\frac12\LCsymb_{\lambda\nu\alpha\beta}\HDT{b\alpha\beta} \itetr{\nu}{\indalg{0}}$, and the 
	fact 
	that $ \partial\itetr{\lambda}{b}/\partial\tetr{a}{\mu} = -\itetr{\lambda}{a}\itetr{\mu}{b} $, 
	we 
	can rewrite \eqref{app.eqn.Noether1} as
	\begin{multline}
		\Laghodge_{\tetr{a}{\mu}} = \LagBE_{\tetr{a}{\mu}} 
		+ \LagBE_{\BT{b}{\lambda}} \frac{\partial \BT{b}{\lambda}}{\partial \tetr{a}{\mu}}
		+ \LagBE_{\ET{b}{\lambda}} \frac{\partial \ET{b}{\lambda}}{\partial \tetr{a}{\mu}} = \\
		\LagBE_{\tetr{a}{\mu}} - \LagBE_{\BT{b}{\lambda}} \KD{\indalg{0}}{a}
		(u^\lambda \BT{b}{\mu} - u^\mu \BT{b}{\lambda}+\LCsymb^{\lambda\mu\alpha\beta} u_\alpha  
		\ET{b}{\beta}) \\
		-\LagBE_{\ET{b}{\lambda}} (u_\lambda \ET{b}{\nu} - u_\nu \ET{b}{\lambda} - 
		\LCsymb_{\lambda\nu\alpha\rho}u^\alpha\BT{b}{\rho})\itetr{\nu}{a}u^\mu.
	\end{multline}
	Using the definitions \eqref{eqn.Legendre2} and \eqref{eqn.Legendre3}, the latter can be 
	rewritten 
	as
	\begin{align}
		&\Laghodge_{\tetr{a}{\mu}} =
		-\Um_{\tetr{a}{\mu}} \nonumber\\
		&- \Um_{\Bm{b}{\lambda}} \KD{\indalg{0}}{a}
		(-u^\lambda \Bm{b}{\mu} + u^\mu \Bm{b}{\lambda}+\LCsymb^{\lambda\mu\alpha\beta} u_\alpha 
		\Um_{\Dm{b}{\beta}}) \nonumber\\
		&-\Dm{b}{\lambda} \itetr{\nu}{a}u^\mu (u_\lambda \Um_{\Dm{b}{\nu}} - u_\nu 
		\Um_{\Dm{b}{\lambda}} + 
		\LCsymb_{\lambda\nu\alpha\rho}u^\alpha\Bm{b}{\rho}),
	\end{align}
	which, after some term rearrangements, reads
	\begin{align}\label{eqn.J.BD}
		&\Laghodge_{\tetr{a}{\mu}} =
		-\Um_{\tetr{a}{\mu}} 
		\nonumber\\
		&+ \KD{\indalg{0}}{a}
		\bigg( 
		u^\lambda \Bm{b}{\mu} \Um_{\Bm{b}{\lambda}} 
		- u^\mu \Bm{b}{\lambda} \Um_{\Bm{b}{\lambda}} 
		\nonumber\\
		&- u^\mu \Dm{b}{\lambda} \Um_{\Dm{b}{\lambda}}
		+ \LCsymb^{\mu\lambda\rho\sigma} u_\rho \Um_{\Bm{b}{\lambda}}
		\Um_{\Dm{b}{\sigma}} 
		\bigg) \nonumber\\
		&- \itetr{\nu}{a}u^\mu
		\left(
		u_\lambda \Dm{b}{\lambda} \Um_{\Dm{b}{\nu}} 
		- \LCsymb_{\nu\lambda\rho\sigma}u^\rho\Bm{b}{\sigma}\Dm{b}{\lambda}
		\right),
	\end{align}
	and exactly is \eqref{eqn.JiA}.
		
	This formula, in particular, can be used to get the following expression for the 
	energy-momentum $ 
	\EMmat{\mu}{\nu} = \tetr{a}{\nu} \NC{a}{\mu}$:
	\begin{align}
		\EMmat{\mu}{\nu} =
		& - \tetr{a}{\nu} \Um_{\tetr{a}{\mu}} \nonumber\\
		& - u^\lambda u_\nu \Bm{a}{\mu} \Um_{\Bm{a}{\lambda}} - u^\mu u_\lambda \Dm{a}{\lambda} 
		\Um_{\Dm{a}{\nu}}				\nonumber\\
		& + u^\mu u_\nu \Bm{a}{\lambda} \Um_{\Bm{a}{\lambda}} 
		+ u^\mu u_\nu \Dm{a}{\lambda} \Um_{\Dm{a}{\lambda}}
		\nonumber \\
		& + \LCsymb_{\nu\sigma\lambda\rho} u^\mu u^\sigma \Bm{a}{\lambda} \Dm{a}{\rho} 
		\nonumber \\
		&+ \LCsymb^{\mu\sigma\lambda\rho} u_\nu u_\sigma \Um_{\Bm{a}{\lambda}} 
		\Um_{\Dm{a}{\rho}}.\label{eqn.sigma.tetr.part}
	\end{align}

	\section{Expression for the energy-momentum}\label{app.energymomentum}
	
	In this section, we derive expression \eqref{eqn.sigma.BD} for the gravitational part fof the 
	energy-momentum
	\begin{equation}\label{eqn.EM.hodge}
		\EMmat{\mu}{\nu} =
		2 \HDT{a\lambda\mu}L_{\HDT{a\lambda\nu}} - 
		(\HDT{a\lambda\rho}L_{\HDT{a\lambda\rho}} - L) \KD{\mu}{\nu}
	\end{equation}
	
	Because $ \HDT{a\mu\nu} $ is 
	antisymmetric tensor, to compute the derivative $ \Laghodge_{\HDT{a\lambda\nu}} $ one needs to 
	use 
	its parametrization via the \We\ connection, which is not symmetric, i.e. $ \HDT{a\mu\nu} = 
	\LCsymb^{\mu\nu\rho\sigma} \w{a}{\rho\sigma}$. Thus, for Lagrangians $ 
	\Lag(\tetr{a}{\mu},\w{a}{\mu\nu}) = 
	\Laghodge(\tetr{a}{\mu},\HDT{a\mu\nu})$ one can write 
	\begin{equation}
		\Lag_{\w{a}{\lambda\mu}} = \LCsymb^{\lambda\mu\rho\sigma} \Laghodge_{\HDT{a\rho\sigma}},
	\end{equation}
	or using the identity $ \LCsymb_{\lambda\mu\alpha\beta}\LCsymb^{\lambda\mu\rho\sigma} = 
	-2(\KD{\rho}{\alpha}\KD{\sigma}{\beta} - \KD{\rho}{\beta}\KD{\sigma}{\alpha}) $, 
	\begin{equation}\label{eqn.L.T}
		\LCsymb_{\alpha\beta\rho\sigma}\Lag_{\w{a}{\lambda\mu}} = -4 \Laghodge_{\HDT{a\rho\sigma}}.
	\end{equation}
	
	On the other hand, using the definitions $ \ET{a}{\mu} = (\w{a}{\mu\nu} - \w{a}{\nu\mu})u^\nu 
	$, $ 
	\BT{a}{\mu} = \LCsymb^{\mu\nu\rho\sigma}\w{a}{\rho\sigma} u_\nu $, and  the parametrization $ 
	\Lag(\tetr{a}{\mu},\w{a}{\mu\nu}) = \LagBE(\tetr{a}{\mu},\BT{a}{\mu},\ET{a}{\mu}) $, one can 
	write
	\begin{align}\label{eqn.Lambda.W}
		\Lag_{\w{b}{\lambda\gamma}} &= 
		\LagBE_{\ET{a}{\mu}} \frac{\partial \ET{a}{\mu}}{\partial \w{b}{\lambda\gamma}} 
		+
		\LagBE_{\BT{a}{\mu}} \frac{\partial \BT{a}{\mu}}{\partial \w{b}{\lambda\gamma}} 
		\nonumber\\
		&=
		u^\gamma \LagBE_{\ET{b}{\lambda}} - u^\lambda \LagBE_{\ET{b}{\gamma}} - 
		\LCsymb^{\lambda\gamma\nu\mu} u_\nu \LagBE_{\BT{a}{\mu}},
	\end{align}
	and hence, from \eqref{eqn.L.T} and \eqref{eqn.Lambda.W}, one can deduce 
	\begin{align}
		\Laghodge_{\HDT{a\lambda\nu}} &= \nonumber\\
		&-\frac{1}{2} 
		\left(
		u_\lambda \LagBE_{\BT{a}{\nu}} - u_\nu \LagBE_{\BT{a}{\lambda}} 
		-
		\LCsymb_{\lambda\nu\rho\sigma} u^\rho \LagBE_{\ET{a}{\sigma}} 
		\right).
	\end{align}
	Then, after contracting the later equation with $ \HDT{a\lambda\mu} = u^\lambda \BT{a}{\mu} - 
	u^\mu 
	\BT{a}{\lambda} + \LCsymb^{\lambda\mu\alpha\beta} u_\alpha \ET{a}{\beta} $, one obtains
	\begin{align}\label{eqn.T.L.T} 
		2 \HDT{a\lambda\mu} &\Laghodge_{\HDT{a\lambda\nu}}  = \nonumber\\
		&  \BT{a}{\mu}\LagBE_{\BT{a}{\nu}} 
		- \ET{a}{\nu}\LagBE_{\ET{a}{\mu}} \nonumber\\
		& + u^\lambda u_\nu \BT{a}{\mu} \LagBE_{\BT{a}{\lambda}} - u^\mu u_\nu \BT{a}{\lambda} 
		\LagBE_{\BT{a}{\lambda}}				\nonumber\\
		& - \LCsymb_{\nu\lambda\rho\sigma} u^\mu u^\sigma \BT{a}{\lambda} \LagBE_{\ET{a}{\rho}} 
		\nonumber\\
		& - \LCsymb^{\mu\lambda\rho\sigma} u_\nu u_\rho \ET{a}{\sigma} \LagBE_{\BT{a}{\lambda}}
		\nonumber \\
		& + (u^\mu u_\nu + \KD{\mu}{\nu}) \ET{a}{\lambda} \LagBE_{\ET{a}{\lambda}} - u^\mu 
		u_\lambda 
		\ET{a}{\nu} \LagBE_{\ET{a}{\lambda}}.
	\end{align}
	This can be used to demonstrate that the full contraction  $ 	\HDT{a\lambda\rho} 
	\Laghodge_{\HDT{a\lambda\rho}} $ results in
	\begin{equation}\label{eqn.T.L.T.full}
		\HDT{a\lambda\rho} \Laghodge_{\HDT{a\lambda\rho}}
		=
		\BT{a}{\lambda} \LagBE_{\BT{a}{\lambda}} + \ET{a}{\lambda} \LagBE_{\ET{a}{\lambda}}.
	\end{equation}
	
	Collecting together \eqref{eqn.T.L.T} and \eqref{eqn.T.L.T.full} and using the change of 
	variables and potential \eqref{eqn.Legendre1}--\eqref{eqn.Legendre3}, we arrive at
	\begin{align}\label{eqn.Sigma.tors.part}
		2 \HDT{a\lambda\mu}&L_{\HDT{a\lambda\nu}} - 
		(\HDT{a\lambda\rho}L_{\HDT{a\lambda\rho}} - L) \KD{\mu}{\nu} = \nonumber\\
		& - \Bm{a}{\mu}\Um_{\Bm{a}{\nu}} - \Dm{a}{\nu}\Um_{\Dm{a}{\mu}} \nonumber\\
		& - u^\lambda u_\nu \Bm{a}{\mu} \Um_{\Bm{a}{\lambda}} - u^\mu u_\lambda \Dm{a}{\lambda} 
		\Um_{\Dm{a}{\nu}}				\nonumber\\
		& + u^\mu u_\nu \Bm{a}{\lambda} \Um_{\Bm{a}{\lambda}} 
		+ u^\mu u_\nu \Dm{a}{\lambda} \Um_{\Dm{a}{\lambda}}
		\nonumber \\
		& + \LCsymb_{\nu\sigma\lambda\rho} u^\mu u^\sigma \Bm{a}{\lambda} \Dm{a}{\rho} 
		\nonumber\\
		& + \LCsymb^{\mu\sigma\lambda\rho} u_\nu u_\sigma \Um_{\Bm{a}{\lambda}} 
		\Um_{\Dm{a}{\rho}} 
		\nonumber \\
		& + (\Bm{a}{\lambda} \Um_{\Bm{a}{\lambda}} + \Dm{a}{\lambda} \Um_{\Dm{a}{\lambda}} - 
		\Um) \KD{\mu}{\nu}.
	\end{align}
	

	\section{Transformation of the torsion PDE}\label{app.sec.Deqn}

	In this appendix, we demonstrate how the Euler-Lagrange equation \eqref{eqn.1st.order.EL} 
	\begin{equation}\label{eqn.EL.tors}
		\D{\nu}(\LCsymb^{\mu\nu\lambda\rho}\Laghodge_{\HDT{a\lambda\rho}}) 
		=-\Laghodge_{\tetr{a}{\mu}}
	\end{equation}
	can be 
	transformed to the form \eqref{eqn.tors.BE.a}.
	
	Based on the different parametrization of the Lagrangian (we omit for the moment dependence of 
	the 
	Lagrangian on the tetrad field) $ \Lag(\w{a}{\mu}) = 
	\Laghodge(\HDT{a\mu\nu}) = \Lagtors(\Tors{a}{\mu\nu}) $, one can obtain
	\begin{equation}\label{eqn.Lambda_W.Hodge}
		\Lag_{\w{a}{\lambda\mu}} = \LCsymb^{\lambda\mu\rho\sigma} \Laghodge_{\HDT{a\rho\sigma}},
		\quad
		\Lag_{\w{a}{\lambda\mu}} = \Lagtors_{\Tors{a}{\lambda\mu}} - 
		\Lagtors_{\Tors{a}{\mu\lambda}},
	\end{equation}
	from which it follows that the objects
	\begin{equation}
		\HTConj{a\mu\nu} := \Laghodge_{\HDT{a\mu\nu}}, 
		\qquad
		\TorsConj{a}{\mu\nu} := \frac12 \left( \Lagtors_{\Tors{a}{\mu\nu}} - 
		\Lagtors_{\Tors{a}{\nu\mu}} 
		\right)
	\end{equation}
	are Hodge duals of each others:
	\begin{subequations}
		\begin{align}
			\HTConj{a\mu\nu} &:=-\frac12\LCsymb_{\mu\nu\rho\sigma}\TorsConj{a}{\rho\sigma}, 
			\\
			\TorsConj{a}{\mu\nu} &:= \phantom{+}\frac12 
			\LCsymb^{\mu\nu\rho\sigma}\HTConj{a\rho\sigma}.
		\end{align}
	\end{subequations}
	Therefore, one can write the following identities
	\begin{subequations}
		\begin{align}
			\HTConj{a\mu\nu} &= u_\mu \Hbb{a}{\nu} - u_\nu \Hbb{a}{\mu} - 
			\LCsymb_{\mu\nu\rho\sigma}u^\rho \Dbb{a}{\sigma}\label{app.eqn.Deqn1},\\[2mm]
			\TorsConj{a}{\mu\nu} &= u^\mu \Dbb{a}{\nu} - u^\nu \Dbb{a}{\mu} +
			\LCsymb^{\mu\nu\rho\sigma}u_\rho \Hbb{a}{\sigma},\label{eqn.TorsConj}
		\end{align}
	\end{subequations}
	where 
	\begin{equation}\label{app.eqn.DH}
		\Hbb{a}{\mu} := \HTConj{a\mu\nu}u^\nu, 
		\qquad
		\Dbb{a}{\mu} := \TorsConj{a}{\mu\nu}u_\nu,
	\end{equation}
	
	Hence, the Euler-Lagrange equation \eqref{eqn.EL.tors} now reads
	\begin{equation}
		\pd{\nu} \TorsConj{a}{\mu\nu} =-\frac12 \Laghodge_{\tetr{a}{\mu}},
	\end{equation}
	or, according to \eqref{eqn.TorsConj}, it can be written as
	\begin{equation}\label{app.eqn.Deqn2}
		\D{\nu}(u^\mu \Dbb{a}{\nu} - u^\nu \Dbb{a}{\mu} + \LCsymb^{\mu\nu\alpha\beta}
		u_\alpha\Hbb{a}{\beta}) = -\frac12 \Laghodge_{\tetr{a}{\mu}}.
	\end{equation}
	
	It remains to express $ \Dbb{a}{\mu} $ and $ \Hbb{a}{\mu} $ in terms of $ 
	\LagBE(\BT{a}{\mu},\ET{a}{\mu}) $. Thus, using the fact that $ \Lag_{\w{a}{\mu\nu}} = 2 
	\TorsConj{a}{\mu\nu} $ and $ \Lag_{\w{a}{\mu\nu}} = \LCsymb_{\mu\nu\rho\sigma} 
	\HTConj{a\rho\sigma} 
	$,  and the expression \eqref{eqn.Lambda.W}, one can derive that
	\begin{subequations}
		\begin{align}
			\Hbb{a}{\mu} &=-\frac12 \left( \LagBE_{\BT{a}{\mu}} + u^\lambda 
			\LagBE_{\BT{a}{\lambda}} 
			u_\mu 
			\right),
			\\
			\Dbb{a}{\mu} &=-\frac12 \left( \LagBE_{\ET{a}{\mu}} + u_\lambda 
			\LagBE_{\ET{a}{\lambda}} 
			u^\mu 
			\right).
		\end{align}
	\end{subequations}
	
	Finally, plugging these expressions in 
	\eqref{app.eqn.Deqn2}, one obtains the desired result 
	\begin{equation}
		\D{\nu}( u^\mu\LagBE_{\ET{a}{\nu}} - u^\nu \LagBE_{\ET{a}{\mu}} + 
		\LCsymb^{\mu\nu\rho\sigma}u_\rho\LagBE_{\BT{a}{\sigma}}) 
		= \Laghodge_{\tetr{a}{\mu}}.
	\end{equation}

	\section{Some expressions of the torsion scalar}
	
	Denoting the scalars in the right-hand side of \eqref{eqn.tors.scal0} as ($ \Tscal = \Tscal_1 + 
	\Tscal_2 + \Tscal_3 $)
	\begin{subequations}
		\begin{align}\label{eqn.tors.scal}
			\Tscal_1 = & \frac14 \mg{ab}g^{\beta\lambda }g^{\mu\gamma 
			}\Tors{a}{\lambda\gamma} \Tors{b}{\beta\mu },\\[2mm]
			\Tscal_2 = & \frac12 g^{\mu\gamma} \itetr{\lambda}{a} \itetr{\beta}{b} 
			\Tors{a}{\mu\beta} 
			\Tors{b}{\gamma\lambda},\\[2mm]
			\quad
			\Tscal_3 = & -g^{\mu\lambda} \itetr{\rho}{a} \itetr{\gamma}{b} \Tors{a}{\mu\rho} 
			\Tors{b}{\lambda\gamma},	  
		\end{align}
	\end{subequations}
	we can also write 
	\begin{subequations}
		\begin{align}\label{eqn.tors.scal.TQLC}
			\mathcal{T}_1 &= Q_3 + \frac12 C_4, \\
			\mathcal{T}_2 &=- Q_1 + Q_4 + L_1 -C_1 + C_2 + \frac12 C_4, \\[2mm]
			\mathcal{T}_3 &= 2 Q_1 + Q_2 + L_2 + 2 C_1 + C_3 - C_4,
		\end{align}
	\end{subequations}
	where the scalars $ Q $, $ L $, and $ C $ are scalars made of $ \ET{a}{\mu} $ and 
	$ \BT{a}{\mu} $ as follows.
	
	\begin{subequations}
	Quadratic in $ \ET{a}{\mu} $:
	\begin{align}\label{eqn.tors.scal2.Q}
		Q_1 &= -\frac12 \ET{\indalg{0}}{\alpha} g^{\alpha\beta} \ET{\indalg{0}}{\beta},\\[2mm]
		Q_2 &= \itetr{\alpha}{a} \ET{a}{\alpha} \itetr{\beta}{b} \ET{b}{\beta} = 
		\ET{\alpha}{\alpha} \ET{\beta}{\beta},\\[2mm]
		Q_3 &=-\frac12 \mg{ab} \ET{a}{\alpha} g^{\alpha\mu}
		\ET{b}{\mu},\\[2mm]
		Q_4 &=-\frac12 \itetr{\lambda}{a} \ET{a}{\beta} \itetr{\beta}{b} \ET{b}{\lambda} = 
		\ET{\lambda}{\beta} \ET{\beta}{\lambda}.
	\end{align}
	
	Mixed scalars (linear in $ \ET{a}{\mu} $):
	\begin{align}\label{eqb.ts.scal.L}
		L_1 &= \LCsymb_{\lambda\tau\varphi\gamma} u^\varphi g^{\tau\beta} \itetr{\lambda}{a} 
		\ET{a}{\beta} \BT{\indalg{0}}{\gamma},\\[2mm]
		L_2 &= 2 \LCsymb_{\lambda\tau\varphi\gamma} u^\varphi g^{\lambda\beta} \itetr{\tau}{a} 
		\ET{\indalg{0}}{\beta} \BT{a}{\gamma}.
	\end{align}
	
	Quadratic in $ \BT{a}{\mu} $ (constant in  $ \ET{a}{\mu} $)
	\begin{align}\label{eqn.tors.scal.C}
		C_1 &=-\frac12 h^{-2} g_{\lambda\rho} \BT{\indalg{0}}{\rho} \BT{\indalg{0}}{\lambda},\\[2mm]
		C_2 &=-\frac12 h^{-2} g_{\rho\varphi} \itetr{\varphi}{a} \BT{a}{\rho} \itetr{\beta}{b} 
		\BT{b}{\lambda}g_{\lambda\beta} 
		,\\[2mm]
		C_3 &= h^{-2} \itetr{\varphi}{a} \BT{a}{\sigma} g_{\varphi\beta} \itetr{\lambda}{b} 
		\BT{b}{\beta} g_{\lambda\sigma} 
		,\\[2mm]
		C_4 &= h^{-2} \itetr{\varphi}{a} \BT{a}{\sigma} g_{\varphi\lambda} \itetr{\lambda}{b} 
		\BT{b}{\beta} g_{\beta\sigma} .
	\end{align}
	\end{subequations}
	
	In terms of the Hodge dual $ \HDT{a\mu\nu} $, the torsion scalar $ \Tscal $ can be rewritten as
	\begin{subequations}
	\begin{equation}\label{eqn.tors.scal.hodge}
		\Tscal = \Hscal_1 + \Hscal_2 + \Hscal_3 + \Hscal_4,
	\end{equation}
	where 
		\begin{align}\label{eqn.hodge.scal}
			\Hscal_1 &= \frac{1}{2 h^2} \mg{ac}\mg{bd}\tetr{c}{\lambda}\tetr{d}{\sigma} 
			g_{\tau\rho} 
			\HDT{a\rho\sigma}\HDT{b\lambda\tau},
			\\[2mm]
			\Hscal_2 &= \frac{1}{2 h^2} \mg{ac}\mg{bd}\tetr{c}{\tau}\tetr{d}{\rho} 
			g_{\lambda\sigma} 
			\HDT{a\rho\sigma}\HDT{b\lambda\tau},
			\\[2mm]
			\Hscal_3 &=-\frac{1}{4 h^2} \mg{ac}\mg{bd}\tetr{c}{\sigma}\tetr{d}{\lambda} 
			g_{\tau\rho} 
			\HDT{a\rho\sigma}\HDT{b\lambda\tau},
			\\[2mm] 
			\Hscal_4 &=-\frac{1}{4 h^2} \mg{ac}\mg{bd}\tetr{c}{\rho}\tetr{d}{\tau} 
			g_{\lambda\sigma} 
			\HDT{a\rho\sigma}\HDT{b\lambda\tau}.
		\end{align}
	\end{subequations}
	
	Finally, in terms of the \We\ connection, the torsion scalar reads:
	\begin{subequations}
		\begin{equation}
			\Tscal = \sum_{n=1}^{8}\mathcal{W}_n
		\end{equation}
		\begin{align}
			\mathcal{W}_1 &= \frac12 
			\mg{ab}g^{\beta\lambda}g^{\mu\gamma}\w{a}{\lambda\gamma}\w{b}{\beta\mu},
			\\
			\mathcal{W}_2 &=-\frac12 
			\mg{ab}g^{\beta\lambda}g^{\mu\gamma}\w{a}{\gamma\lambda}\w{b}{\beta\mu},
		\end{align}
		\begin{align}
			\mathcal{W}_3 &= \frac12 
			g^{\mu\gamma}\itetr{\beta}{a}\w{a}{\lambda\gamma}\itetr{\lambda}{b}\w{b}{\beta\mu},
			\\
			\mathcal{W}_4 &=\frac12 
			g^{\mu\gamma}\itetr{\beta}{a}\w{a}{\gamma\lambda}\itetr{\lambda}{b}\w{b}{\mu\beta},
		\end{align}
		\begin{equation}
			\mathcal{W}_5 =  
			-g^{\mu\gamma}\itetr{\beta}{a}\w{a}{\gamma\lambda}\itetr{\lambda}{b}\w{b}{\beta\mu},
		\end{equation}
		\begin{align}
			\mathcal{W}_6 &= -g^{\mu\lambda} w^{1}_{\lambda} w^{1}_{\mu},
			\\
			\mathcal{W}_7 &= -g^{\mu\lambda} w^{2}_{\lambda} w^{2}_{\mu},
			\\
			\mathcal{W}_8 &= 2 g^{\mu\lambda} w^{1}_{\mu} w^{2}_{\lambda},
		\end{align}
		where $ w^{1}_\mu = \itetr{\lambda}{a} \w{a}{\mu\lambda}$, $ w^{2}_\mu = \itetr{\lambda}{a} 
		\w{a}{\lambda\mu} $.
	\end{subequations}

	\printbibliography
	
\end{document}

%% file: structure.tex
\usepackage[
nochapters, 
]{classicthesis} 

\usepackage{arsclassica} 

\usepackage{enumitem} 

\usepackage{amsmath,amssymb,amsthm} 

\usepackage{varioref} 

\usepackage{accents}

\usepackage[title]{appendix}

\usepackage[symbol]{footmisc}

\usepackage{hyperref}

\theoremstyle{definition} 

\theoremstyle{plain} 

\theoremstyle{remark} 

\usepackage[backend=biber,giveninits=true,url=false,doi=true,eprint=true,isbn=false,
backref,backrefstyle=none,sorting=nyt,maxbibnames=99]{biblatex}
\DefineBibliographyStrings{english}{%
  backrefpage = {Cited on p\adddot},%
  backrefpages = {Cited on pp\adddot}%
}

\bibliography{biblio}

\renewbibmacro{in:}{%
  \ifboolexpr{%
     test {\ifentrytype{article}}%
  }{}{\printtext{\bibstring{in}\intitlepunct}}%
}

\usepackage{empheq}
\newlength\mytemplen
\newsavebox\mytempbox
\makeatletter
\definecolor{cream}{rgb}{.81, .88, 1}
 \newcommand\mycreambox{%
     \@ifnextchar[
        {\@mycreambox}%
        {\@mycreambox[0pt]}}
 \def\@mycreambox[#1]{%
     \@ifnextchar[
        {\@@mycreambox[#1]}%
        {\@@mycreambox[#1][0pt]}}
 \def\@@mycreambox[#1][#2]#3{
     \sbox\mytempbox{#3}%
     \mytemplen\ht\mytempbox
     \advance\mytemplen #1\relax
     \ht\mytempbox\mytemplen
     \mytemplen\dp\mytempbox
     \advance\mytemplen #2\relax
     \dp\mytempbox\mytemplen
     \colorbox{cream}{\hspace{1em}\usebox{\mytempbox}\hspace{1em}}}
 \makeatother
 
 \hypersetup{
 	colorlinks=true, breaklinks=true, bookmarks=true,bookmarksnumbered,
 	urlcolor=webbrown, linkcolor=RoyalBlue, citecolor=webgreen, 
 	pdftitle={}, 
 	pdfauthor={\textcopyright}, 
 	pdfsubject={}, 
 	pdfkeywords={}, 
 	pdfcreator={pdfLaTeX}, 
 	pdfproducer={LaTeX with hyperref and ClassicThesis} 
 }